\documentclass[sigconf,10pt]{acmart}

\setcopyright{none}
\renewcommand\footnotetextcopyrightpermission[1]{}
\usepackage{hyperref}
\usepackage{multirow}
\usepackage{graphicx}
\usepackage{subfigure}
\usepackage[utf8]{inputenc}
\usepackage[ruled, lined, longend, linesnumbered]{algorithm2e}
\usepackage{amsmath,multicol}

\usepackage{amssymb,textcomp,comment}
\usepackage{makecell}
\usepackage{ulem}

\usepackage{booktabs}
\usepackage{multirow}
\usepackage{color}
\usepackage{colortbl}

\usepackage{listings}
\usepackage{enumitem}
\usepackage{cleveref}

\crefname{section}{§}{§§}
\Crefname{section}{§}{§§}

\definecolor{dkgreen}{rgb}{0,0.6,0}
\definecolor{gray}{rgb}{0.5,0.5,0.5}
\definecolor{mauve}{rgb}{0.58,0,0.82}

\usepackage{array}
\newcolumntype{L}[1]{>{\raggedright\let\newline\\\arraybackslash\hspace{0pt}}m{#1}}
\newcolumntype{C}[1]{>{\centering\let\newline\\\arraybackslash\hspace{0pt}}m{#1}}
\newcolumntype{R}[1]{>{\raggedleft\let\newline\\\arraybackslash\hspace{0pt}}m{#1}}

\newcommand{\mwx}[1]{\textbf{\color{red}{#1}}}
\newcommand{\xdl}[1]{\textbf{\color{blue}{#1}}}
\newcommand{\sys}{\texttt{Mandheling}\xspace}

\newcommand{\xzl}[1]{\textbf{\color{violet}{(Xuanzhe: #1)}}}

\begin{document}

	\title{\sys: Mixed-Precision On-Device DNN  Training with DSP Offloading}
	
	\author{Daliang Xu$^{1*}$, Mengwei Xu$^{2*}$$^\#$, Qipeng Wang$^{1}$, Shangguang Wang$^2$, Yun Ma$^1$, Kang Huang$^3$, Gang Huang$^1$, Xin Jin$^1$, Xuanzhe Liu$^{1\#}$}
\affiliation {
	\institution{$^1$Key Lab of High Confidence Software Technologies (Peking University), Beijing, China}
	\country{}
}
\affiliation {
	\institution{$^2$State Key Laboratory of Networking and Switching Technology (BUPT), Beijing, China}
	\country{}
}
\affiliation {
	\institution{$^3$Linggui Tech Company, Beijing, China.}
	\country{}
}
\thanks{*Equal contributions; $^\#$Corresponding authors.}

\renewcommand{\shortauthors}{Daliang Xu et al.}

	\begin{abstract}

This paper proposes \sys, the first system that enables highly resource-efficient on-device training by orchestrating the mixed-precision training with on-chip Digital Signal Processing (DSP) offloading.
\sys fully explores the advantages of DSP in integer-based numerical calculation by four novel techniques:
(1) a CPU-DSP co-scheduling scheme to mitigate the overhead from DSP-unfriendly operators;
(2) a self-adaptive rescaling algorithm to reduce the overhead of dynamic rescaling in backward propagation;
(3) a batch-splitting algorithm to improve the DSP cache efficiency;
(4) a DSP-compute subgraph reusing mechanism to eliminate the preparation overhead on DSP.
We have fully implemented \sys and demonstrate its effectiveness through extensive experiments.
The results show that, compared to the state-of-the-art DNN engines from \texttt{TFLite} and \texttt{MNN}, \sys reduces the per-batch training time by 5.5$\times$ and the energy consumption by 8.9$\times$ on average.
In end-to-end training tasks, \sys reduces up to 10.7$\times$ convergence time and 13.1$\times$ energy consumption, with only 1.9\%--2.7\% accuracy loss compared to the FP32 precision setting.
\end{abstract}

	\maketitle














	\section{Introduction}\label{sec:intro}

With the ever-increasing concerns of privacy~\cite{gdpr}, empowering a mobile device to train a deep neural network (DNN) locally, i.e., \textit{on-device training} is recently attracting attentions from both academia and industry~\cite{deeptype,siri-fl,gboard-fl}.
Without giving away the training data, on-device training enables
(i) geo-distributed devices to collaboratively establish a high-accuracy model~\cite{hard2018federated,bonawitz2019towards}
or
(ii) a single device to personalize and adapt its model to its own environments~\cite{deeptype}.

However, a key obstacle towards practical on-device training is its huge resource cost.
According to our preliminary measurements with two popular DL libraries (\texttt{TFLite}~\cite{tflite} and \texttt{MNN}~\cite{mnn}), training ResNet-50 with one batch (batch size 32) takes 4.6 GB memory and 36.4 seconds on the smartphone XiaoMI 10 equipped with Snapdragon 865 CPU.
The consumed energy equals to watching a 1080P-definition video for 111.2 seconds.
In an end-to-end learning scenario, it typically takes thousands or even more such batches of training and the accumulated cost becomes prohibitively expensive.
Unfortunately, this issue has not been well explored by the research community.
Existing studies gaining impressive benefits for on-device inference tasks~\cite{wang2021asymo,chen2018tvm,liu2019optimizing,huynh2017deepmon,xu2018deepcache} can be hardly applied to on-device training due to the huge gap between inference and training workloads, e.g., the different computation patterns and accuracy requirements.

To optimize the performance of on-device training, this work is motivated by two key observations.
First, traditional DNN training mostly  performs on FP32 data format to achieve good model accuracy.
However, the ML community has recently proposed various \textit{mixed-precision training} algorithms~\cite{wang2020niti,zhou2021octo,wu2016binarized,lin2015neural,rastegari2016xnor, zhou2018adaptive,lin2017towards,courbariaux2015binaryconnect,jacob2018quantization,zhang2020fixed,yang2020training,zhong2020towards}, where the weights and activations generated during training are represented not only by FP32 but also by  lower-precision formats such as INT8 and INT16.
By exploiting the hardware features in accelerating integer operations, these algorithms have been demonstrated to be effective in reducing the training-time resource cost while guaranteeing the convergence accuracy, i.e., only 1.3\% loss on CIFAR-10~\cite{yang2020training}.
Second, modern mobile SoCs often consist of heterogeneous processors, among which the Digital Signal Processor (DSP) is ubiquitously available and particularly suits integer operations, i.e., INT8-based matrix multiplication.
For example, Hexagon 698 DSP~\cite{QualcommHexagon} is adequate to execute 128 INT8 operations in one cycle and has been demonstrated to be 11.3$\times$/4.0$\times$ more energy-efficient than CPU/GPU in DL inference tasks~\cite{zhang2022comprehensive}, respectively.
Intuitively, under the mixed-precision training setting, we are interested in a question whether we can partially offload the training workloads, especially those integer-based operations, from CPU to DSP to reduce the cost of on-device training.

In this paper, we propose a first-of-its-kind system, namely \sys, which enables highly resource-efficient, mixed-precision on-device training with on-chip DSP offloading.
To facilitate developers in using different types of training algorithms on \sys, we investigate popular mixed-precision training algorithms and extract the key principles from them.
Based on those principles, we incorporate a unified abstraction into \sys as will be discussed in $\S$\ref{sec:design:abstraction}.
With a given training algorithm and the model to be trained, \sys aims to minimize the training cost by judiciously co-scheduling various training operators to mobile DSP (mostly) and CPU.
When designing \sys, however, we need to address the following major challenges that have not been explored in prior literature.
\begin{itemize}[leftmargin=0pt,itemindent=10pt,topsep=0pt]
	\item \textit{Dealing with DSP-unfriendly operators.}
	In a typical mixed-precision training algorithm, some operators like Transpose and Normalization run slowly on DSP due to their irregular memory accesses~\cite{mittal2016survey} or lack of architecture-level support on DSP.
	A judicious scheme to determine what operators to be offloaded to DSP while others are placed on CPU is needed to fully exploit those heterogeneous processors.
	\item \textit{Dynamic rescaling does not fit DSP.}
	During the training, dynamic rescaling~\cite{wang2020niti,zhang2020fixed,yang2020training,zhong2020towards} is a critical operation to quantize/dequantize among different data types.
	We observe that this operation runs slowly on DSP and can easily compromise the benefits of using DSP.
	Because dynamic scaling is inserted into each layer with the low-precision data type, simply scheduling it to CPU like other DSP-unfriendly operators incurs high context switching overhead, therefore needs to be optimized exclusively.
	\item \textit{Exhausted data cache.}
	Training workloads impose high pressure on the DSP cache and a vanilla implementation leads to a  low cache hit ratio.
	The reasons are twofold.
	First, training tasks often require a large batch size, which results in frequent memory accesses for accessing  intermediate data.
	Second, the DSP cache is often smaller than the CPU cache, e.g., only half for L2 cache on Snapdragon 865.
	Considering that fully utilizing the processor cache is a killing factor towards memory-intensive operations such as convolution weight gradients~\cite{wang2021asymo}, exhausted data cache on DSP is likely to act as the bottleneck in the training process.
	\item \textit{Costly compute graph preparation.}
	Unlike inference tasks, on-device training tasks  usually use dynamic graphs to facilitate developers in development and debugging~\cite{pytorch}.
	However, preparing the compute graph on DSP takes a considerable amount of time for allocating DSP memory, building graph-to-operation reference, etc.
	Therefore, how to eliminate the compute graph preparation under a tight memory budget of DSP is critical to \sys.
\end{itemize}

\textbf{Key techniques of \sys.}
To address the preceding challenges, \sys presents the following novel techniques to fully unleash the DSP computing capacity:
\begin{itemize}[leftmargin=0pt,itemindent=10pt,topsep=0pt]
\item \textbf{CPU-DSP co-scheduling} \noindent ($\S$\ref{sec:parallel}) is proposed to mitigate the overhead of DSP-unfriendly operators.
The key idea is to reduce the number of context switchings brought by DSP-unfriendly operators and overlap the CPU and DSP execution as much as possible.
This is achieved through our novel scheduling algorithm that considers the latency of the operator executing on different processors and the overhead of CPU-DSP context switching.

\item \textbf{Self-adaptive rescaling} \noindent ($\S$\ref{sec:rescaling}) is proposed to significantly reduce the overhead of dynamic rescaling by adaptively lowering its invoking frequency.
This is motivated by our micro-experiments which demonstrate that, after the early stage of training, the actual changing frequency of the scale factor becomes low and its value becomes fairly stable.

\item \textbf{Batch splitting} \noindent ($\S$\ref{sec:splitting})  is proposed to reduce the cache pressure on DSP and so as to increase the cache hit ratio.
\sys runs the intra-operator partition at the batch dimension because this solution does not influence the inputs and weights of the original convolution operation, thus causing no redundant computations.
\sys uses an intuitive yet effective method to identify the splitting point of batch size.
It also provides an integer-only scheme to efficiently concatenate the output from the split batch.

\item \textbf{DSP-compute subgraph reuse} \noindent ($\S$\ref{sec:reuse}) is proposed to eliminate the preparation overhead of DSP compute graph.
This is motivated by a key observation that the model structure is rarely changed during the training phase.
To further tackle the memory constraint of DSP, we provide a practical subgraph-reusing algorithm based on the minimum dynamic memory allocation/deallocation principle.
\end{itemize}

\textbf{Implementation and evaluation}
We have fully implemented \sys with 15k LoC in C/C++ and 800 LoC in assembly language.
\sys is a standalone framework that supports models exported from different front-end frameworks, e.g., TensorFlow~\cite{tensorflow} and PyTorch~\cite{pytorch} and is compatible with various mixed-precision training algorithms.
We then conducted extensive experiments on 6 typical DNN models (VGG-11/16/19~\cite{simonyan2014very}, ResNet-18/34~\cite{he2016deep}, and InceptionV3~\cite{szegedy2016rethinking}) and 3 commodity mobile devices (XiaoMI 11 Pro, XiaoMI 10, and Redmi Note9 Pro).
The results demonstrated that, compared to native supports from \texttt{TFLite} and \texttt{MNN}, \sys can reduce the per-batch training time and energy consumption by 5.5$/$8.9$\times$ on average and up to 8.3$\times$/12.5$\times$, respectively.
Furthermore, compared to GPU-enhanced training, \sys speeds up the per-batch training time by 7.1$\times$ and reduces the energy consumption by 5.8$\times$ on average.
In an end-to-end training on a single device, compared with FP32-based training algorithm and \texttt{MNN}, \sys accelerates the model convergence by 5.7$\times$ on average and reduces the total energy consumption by 7.8$\times$ on average.
The improvements are even more profound in a federated learning scenario with 8.0$\times$ convergence speedup and 10.6$\times$ energy reduction on average.
Meanwhile, the accuracy of trained models drops marginally with only 1.9\%-2.7\% loss compared to the accuracy of FP32 precision setting, which is consistent with the theoretical  results adopted by the ML community~\cite{wang2020niti,zhou2021octo,wu2016binarized,lin2015neural,rastegari2016xnor, zhou2018adaptive,lin2017towards,courbariaux2015binaryconnect,jacob2018quantization,zhou2016dorefa}.
The ablation study further distinguishes  the effectiveness of every single key technique  of \sys.

\textbf{Contributions} are summarized as following.
\begin{itemize}[leftmargin=10pt,topsep=0pt]
	\item We thoroughly explore the opportunities and challenges of DSP offloading for mixed-precision on-device training. 
	\item We design and implement the first DSP-offloading based mixed-precision on-device training framework, which incorporates four novel techniques, i.e., self-adaptive rescaling, batch splitting, CPU-DSP co-scheduling, and DSP compute subgraph resue.
	The system will be fully open-sourced soon on Github.
	\item We evaluate \sys with representative DNN models and commodity mobile devices. The results demonstrate \sys's superior effectiveness and practical value.
\end{itemize}

	\section{Background and Motvations}\label{sec:bkgnd}
We briefly introduce some background and our motivations.

\subsection{On-Device DNN Training}

A trend to deploy DNN training on devices locally is emerging, especially with the data privacy concerns (like GDPR \cite{gdpr}) in AI applications. 
Similar to DNN training on the cloud/server, the on-device training also conducts mini-batch sampling strategy, where each batch's training comprises of three stages: the forward pass, the backward pass, and the weight update.
The forward pass loads inputs and calculates the loss;
the backward pass usually employs a specific optimizer like Stochastic Gradient Descent (SGD) \cite{bottou2010large} to obtain the gradients;
lastly, the gradients are applied to the weights for model update.
Compared to the on-device inference, the training task is more resource-consuming because:
(i) the backward pass contains about 2$\times$ FLOPs as forward pass (the same as inference);
(ii) a training process often involves hundreds or even thousands of mini-batches.

\begin{figure}[t]
	\centering
	\subfigure[Comparison of inference and training time (batch size = 64.)] {\includegraphics[width=0.22\textwidth]{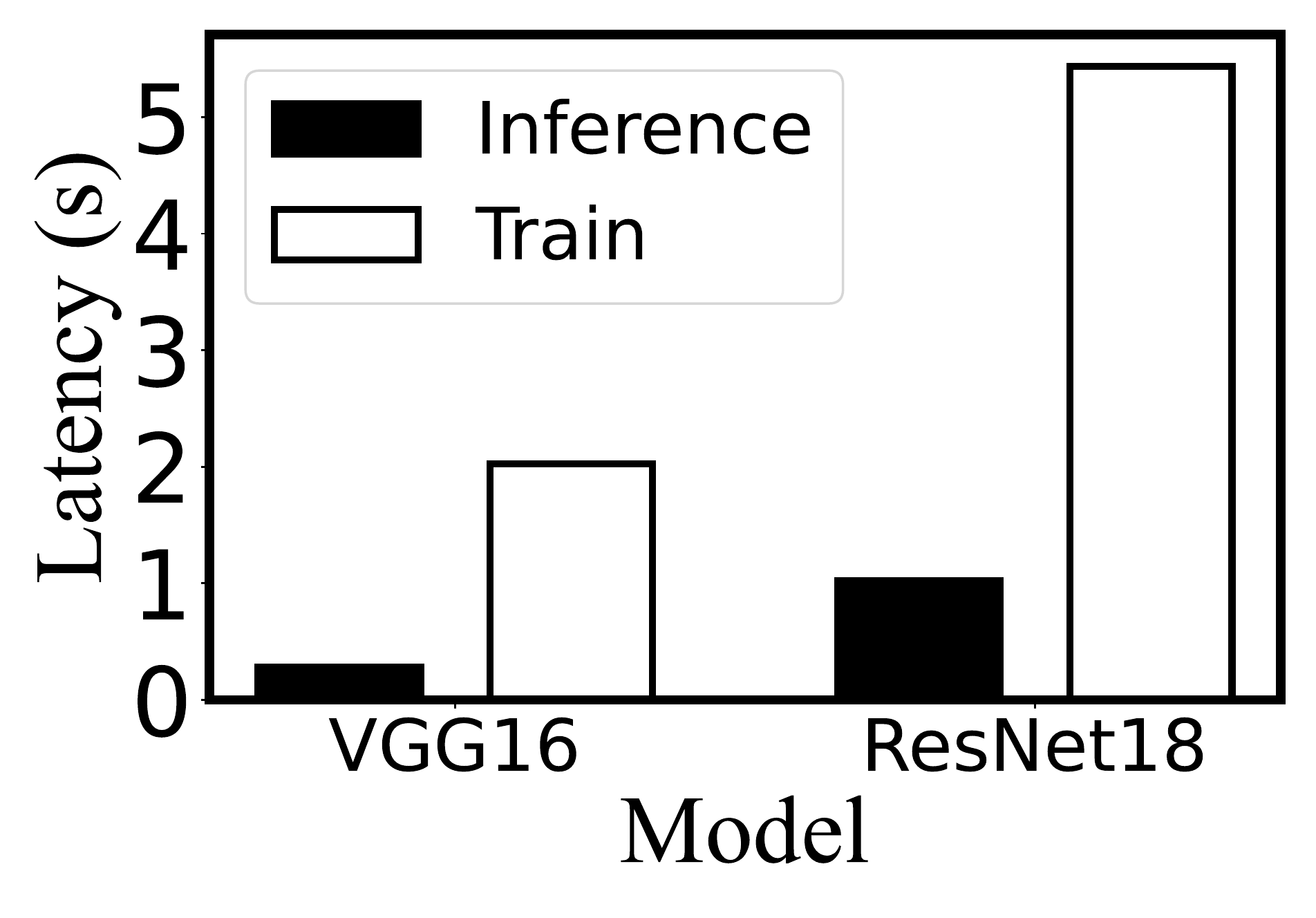} } 
	\subfigure[Power consumption of different Apps and DNN training on devices.]{\includegraphics[width= 0.24\textwidth ]{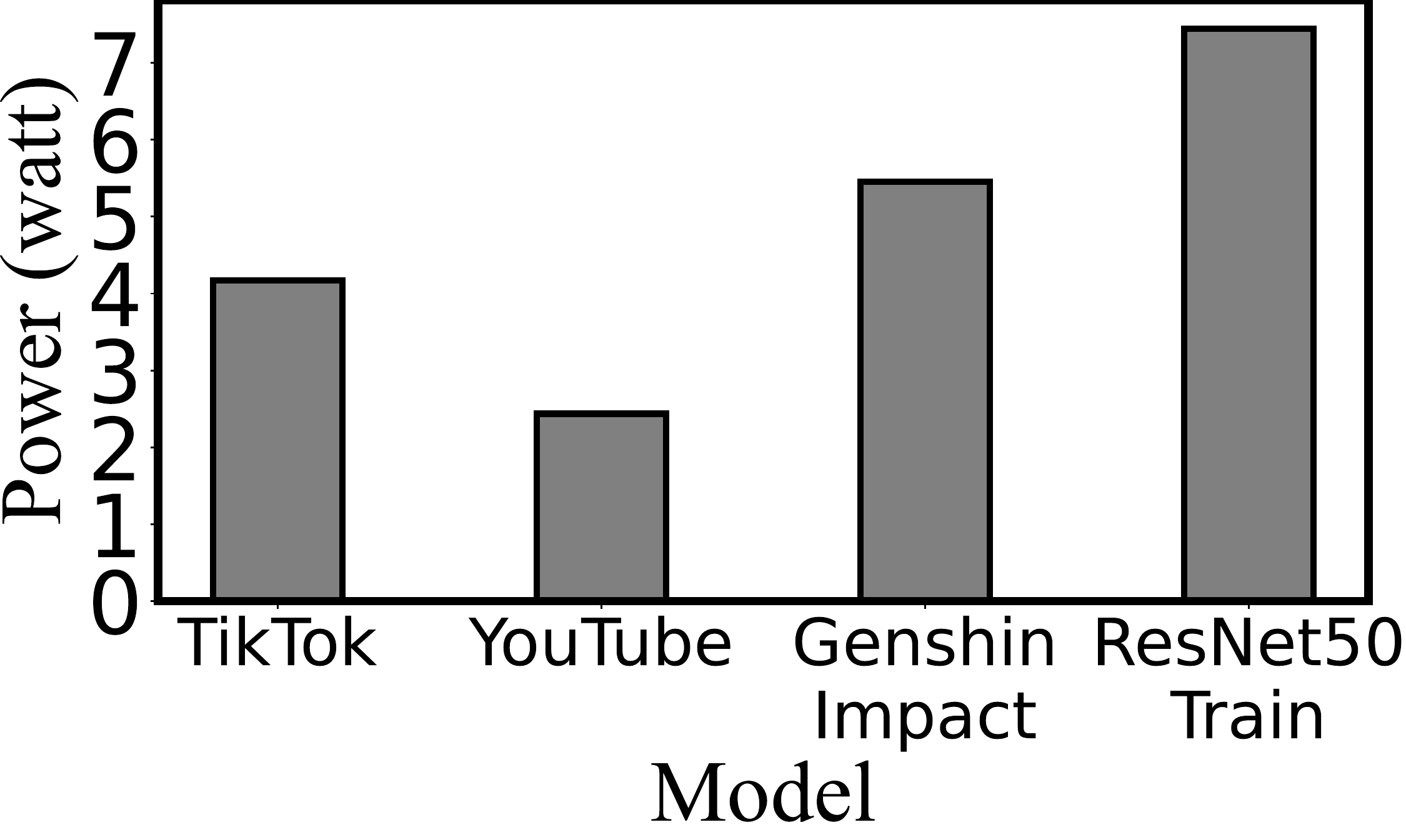} }
	\vspace{-0.5cm}
			\caption{Preliminary measurements that highlight the huge resource cost of on-device training.} 
				\label{fig-metrics}
\vspace{-0.4cm}
\end{figure}
We conducted two measurement studies to highlight the prohibitively high resource cost of on-device training.
Figure \ref{fig-metrics}(a) shows that the DNN training takes 7.14$\times$ more time than inference with the same batch size, which is larger than the theoretical FLOPs gap (about 3$\times$).
That is because training ops, especially for weight gradient ops, are more difficult to optimize because of the dynamic input width and height of the feature map.
Besides, Figure \ref{fig-metrics}(b) shows that on-device training consumes more energy than typical video players (TikTok~\cite{tiktok}, YouTube~\cite{youtube}) and gaming apps (Genshin Impact \cite{genshin}).
For instance, training one batch (BS=32) of the ResNet-50 model costs the same energy as watching 36.4 seconds of videos on YouTube.
Considering the minimal granularity for training is usually one epoch which, let's say, contains 1,000 mini-batches, the energy cost is as high as watching 7.91 hours of videos on YouTube.

\subsection{Mixed-Precision Training} \label{sec:low-precision-training}

\begin{figure}[t]
	\centering
	 \includegraphics[width=0.45\textwidth]{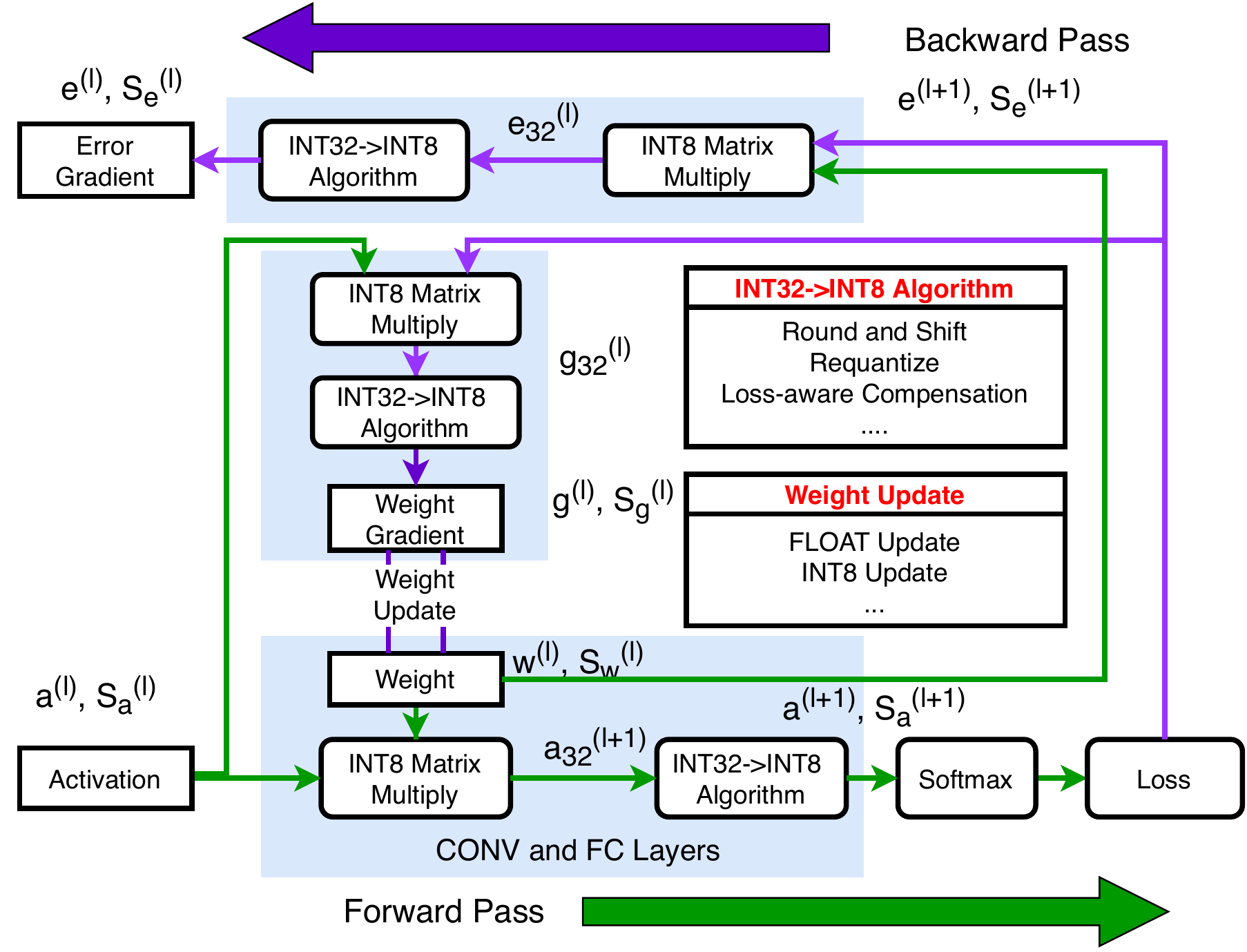}
	 \vspace{-0.4cm}
	\caption{Workflow of mixed-precision training.} 
	\label{fig-int8-training}
	\vspace{-20pt}
\end{figure}

To reduce the resource cost of DNN training, mixed-precision training algorithms have been proposed \cite{wang2020niti,zhou2021octo,wu2016binarized,lin2015neural,rastegari2016xnor, zhou2018adaptive,lin2017towards,courbariaux2015binaryconnect,jacob2018quantization,zhou2016dorefa}.
These algorithms mainly exploit the feature that the data in DNNs (including the activation, weights, bias, and gradients) are usually in high redundancy \cite{louizos2018relaxed}, therefore their representation precision can be reduced, e.g., from FP32 to FP16/INT8.
Fewer bits per number benefits the DNN training to run faster through the single-instruction-multiple-data (SIMD) hardware parallelism, which is commonly available on mobile processors \cite{wang2021comprehensive}.
Here, we use INT8 training as an example to illustrate how typical mixed-precision training algorithms work.
Figure~\ref{fig-int8-training} shows an exemplified workflow.

\noindent\textbf{Forward pass.}
After quantization, activation $a^{(l)}$ and weight $w^{(l)}$ are INT8 numbers with scale factors $S_a^{(l)}$ and  $S_w^{(l)}$.
With INT8 matrix multiply which is used to replace traditionally FP32-based multiplication, we can obtain intermediate INT32 activation $a_{32}^{(l+1)}$.
To transform the intermediate results to INT8 numbers, an INT32-to-INT8 algorithm is needed, such as Round and Shift \cite{wang2020niti}, Loss-aware Compensation \cite{zhou2021octo}, or Requantize \cite{tensorflow}.
When forwarding to the final layer, the activations are input to the softmax and loss layer.

\noindent\textbf{Backward pass and weight update.}
The obtained INT8 error gradients $e^{(l+1)}$ will multiply with $w^{(l)}$ and $a^{(l)}$ to obtain intermediate INT32 error gradients $e_{32}^{(l)}$ and weight gradients  $g_{32}^{(l)}$.
Also through the INT32 to INT8 algorithm, we can get the error and weight gradients of the l-th layer $e^{(l)}$ and  $g^{(l)}$ with their scale factor $S_e^{(l)}$ and $S_g^{(l)}$.
 Finally, the model weights are updated by gradient $g^{(l)}$ with the global learning rate and other hyperparameters.
The update method can be divided into two categories: FLOAT update and INT8 update, which means the former one can support changing scale factor.


\subsection{Mobile DSP Offloading}

DSP (Digital Signal Processor) is originally designed for processing digital signals like audio with high energy efficiency.
Almost each mobile SoC includes DSP, where the most common is Hexagon produced by Qualcomm~\cite{QualcommHexagon}.
In 2016 Qualcomm announced Hexagon 680 DSP -- the first DSP with Hexagon Vector Extensions (HVX) designed to allow significant compute workloads for advanced imaging, and computer vision \cite{dsp-processor}.
Given its popularity, this work targets Hexagon DSP to run the DNN training workloads, but its techniques are compatible with other DSP hardware as well.

\noindent\textbf{Hexagon DSP architecture.}
Nowaday Hexagon DSP contains hexagon cores and a SIMD coprocessor.
The former one performs general-purpose processing while the latter one is good at vector computation \cite{hexagon-dsp-sdk}.
\sys mainly exploits the SIMD co-processor which can process 1024-bit fixed point data inside one HVX instruction, or 128  INT8 mathematical functions like \texttt{add} and \texttt{multiply} in one cycle.
Besides, the hexagon core's clock frequency is 500 MHz which is much lower than the CPU ones so that it is much more energy-friendly.
However,
its hexagon cores are too weak to perform heavy general processing and its SIMD co-processor does not include float processing units to perform FP32/FP16 operations.
Therefore, we need to carefully design the training operations on Hexagon DSP to gain the expected benefits.

\noindent\textbf{Hexagon DSP programming model.}
The Hexagon DSP and CPU cores share the main memory, but do not share the cache. The two processors have their own memory space, indicating that data needs to be copied between them.
Therefore, a typical program contains two parts: the application logic code running on CPU and the data processing code running on DSP.
DSP code is dynamically loaded on invocation of synchronous Remote Procedure Call (FastRPC) \cite{hexagon-dsp-sdk}.

\begin{figure*}[t]
	\centering
	\includegraphics[width=0.95\textwidth]{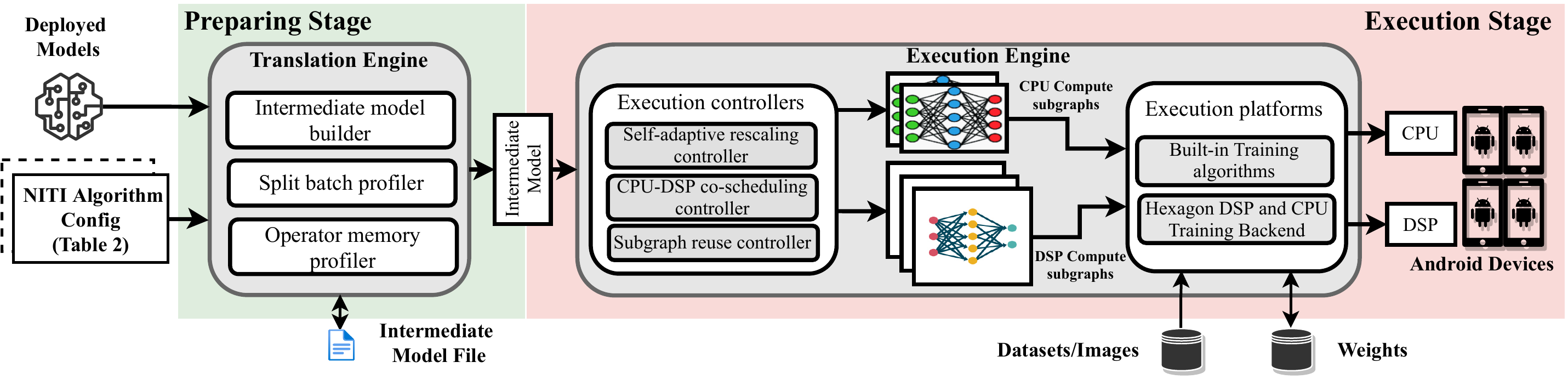}
	\vspace{-0.5cm}
	\caption{The overall workflow of \sys.}
	\label{fig-workflow}
	\vspace{-0.2cm}
\end{figure*}

	\section{The Design}\label{sec:design}

\begin{table*}[t]
	\begin{minipage}[t]{0.55\textwidth}
	\smaller
	\renewcommand\arraystretch{1.22}
	\begin{tabular}{c|cccc|c}
		\hline
		\textbf{Mixed-precision algo.} & \textbf{W} & \textbf{A} & \textbf{G} &  \textbf{WU} & \textbf{support} \\ \hline
		NITI~\cite{wang2020niti} &  INT8 & INT8 & INT8 &  INT8 & $\checkmark$ \\ \hline
		Octo~\cite{zhou2021octo} &  INT8 & INT8 & INT8 &  INT8 & $\checkmark$ \\ \hline
		Adaptive Fixed-Point~\cite{zhang2020fixed} &  INT8/INT16 & INT8 & INT8 &  FP32 & $\checkmark$  \\ \hline
		WAGEUBN~\cite{yang2020training}  &  INT8 & INT8 & INT8 &  FP24 & $\checkmark$ \\ \hline
		MLS Format~\cite{zhong2020towards} &  INT8 & INT8 & INT8 &  FP32 & $\checkmark$ \\ \hline
		Chunk-based~\cite{wang2018training} & FP8 & FP8 & FP8 &  FP16 & $\times$ \\ \hline
		Unified INT8 Training\cite{zhu2020towards} &  INT8 & INT8 & INT8 &   FP32 & $\times$ \\ \hline
		\multicolumn{6}{c}{	"W", "A", "G", and "WU" represent weight, activation, gradient and weight update.}
	\end{tabular}
	\caption{Data types of mixed-precision training algorithms.}
	\label{table-low-precision-algorithms}
	\vspace{-0.7cm}
	\end{minipage}\hfill
	\begin{minipage}[t]{0.43\textwidth}
	\smaller
	\setlength{\tabcolsep}{1mm}{
	\begin{tabular}{c|c|c}
		\hline
		\multirow{2}{*}{\textbf{Attribute}} & \multicolumn{2}{c}{\textbf{Contents}}  \\ \cline{2-3}
		 & \textbf{key}  & \textbf{value} \\ \hline
		\multirow{2}{*}{Translation}  & FP32 Conv & \scriptsize INT8 Conv+ReduceMax+Shift  \\ \cline{2-3}
		 & FP32 MaxPool & INT8 MaxPool \\ \hline
		\multirow{2}{*}{Backprop.} &  FP32 Conv Error Grad. & INT8 Deconv \\ \cline{2-3}
		&   \scriptsize FP32 Conv Weight Grad.&  \scriptsize INT8 ConvBackpropFilter \\ \hline
		\multirow{3}{*}{Weight} & Initializer &  Xavier\_normal\\ \cline{2-3}
		 & Type &  INT8 \\  \cline{2-3}
		 & Update &  INT8 \\  \hline
		\multirow{2}{*}{Optimizer} & Loss & Cross Entropy   \\  \cline{2-3}
		& Optimizer & SGD   \\  \hline
	\end{tabular} }
	\caption{A typical NITI algorithm training config.}
	\label{tab-abstraction}
	\vspace{-0.7cm}
	\end{minipage}
\end{table*}

\subsection{Overview}


\noindent \textbf{Design goal} \sys aims to minimize the latency and energy consumption under a given training task through on-device DSP offloading.
\sys is designed as a generic framework to support different kinds of mixed-precision algorithms and allows users to customize the  algorithms through its exposed abstraction and configurations as will be discussed in $\S$\ref{sec:design:abstraction}.
The convergence accuracy is guaranteed by the used mixed-precision training algorithm.

\noindent \textbf{Workflow.}
Figure~\ref{fig-workflow} illustrates the workflow of \sys.
The input of \sys includes the mixed-precision training algorithms and the model file that can be either pre-trained or randomly initialized.
Once \sys is deployed on a device, it works in two following stages:
(1) At the preparing stage, \sys translates models from different front-end frameworks (e.g., TensorFlow~\cite{tensorflow} and PyTorch~\cite{pytorch}) to intermediate models in form of FlatBuffer-format model file.
(2) At the execution stage, \sys generates CPU and DSP compute subgraphs and performs compute subgraphs execution on Android devices.
Note that both stages run on devices, and the preparation stage is automatically triggered before the first-time execution of one shot.
Hence, such a design does not introduce any additional programming efforts to the app developers.
\begin{itemize}[leftmargin=0pt,itemindent=10pt,topsep=0pt]
	\item \textbf{Preparing stage.}
	When a to-be-trained model is downloaded to a device, \sys translates the model to an intermediate model via the  \textit{Intermediate Model Builder} according to mixed-precision training configuration.
	The intermediate model contains the operator's type, hyperparameters, inputs and outputs as well as the memory regions of intermediate model inputs and outputs.
	Then, \sys runs a profiling iteration to obtain each layer's optimal configuration via \textit{Split Batch Profiler} to further optimize intermediate model to gain higher performance as to be shown in $\S$\ref{sec:splitting}.
	Lastly, a FlatBuffer format model file will be generated to be processed directly by the \sys runtime.

	\item \textbf{Execution stage.}
	Once a training task starts, \sys runtime first loads datasets and weights from disk to the required memory regions, and then generates CPU and DSP compute subgraphs via \textit{CPU-DSP co-scheduling controller} ($\S$\ref{sec:parallel}).
	All subgraphs will execute on  Android devices' CPU or DSP via \sys's \textit{Hexagon DSP and CPU Training backend}, which includes operator implementation optimized for mixed-precision data types on CPU and DSP.
\end{itemize}


\textbf{Key techniques.}
While DSP has been proven to be useful in DNN inference~\cite{lee2019mobisr}, we observe a disparity between using DSP to serve training and inference tasks.
Compared to the CPU, a vanilla DSP training engine can achieve very limited or even negative performance gain.
To this end, \sys incorporates four key techniques to unleash the DSP computing capacity.
The techniques can be divided into two major classes: \textit{intra-operator} and \textit{inter-operator}.
At the intra-op level, since convolution and weight gradient layers will bring excessive memory access and exhaust the DSP cache, we use self-adaptive rescaling ($\S$\ref{sec:rescaling}) and batch splitting ($\S$\ref{sec:splitting}) to reduce memory access and increase the cache hit ratio.
At the inter-op level, \sys overlaps the CPU and DSP execution as much as possible to mitigate the overhead of DSP-unfriendly operators ($\S$\ref{sec:parallel}),
and reuses DSP compute graph to eliminate the preparation overhead ($\S$\ref{sec:reuse}).

\subsection{Mixed-Precision Training Abstraction}\label{sec:design:abstraction}
We thoroughly investigate the typical mixed-precision training algorithms~\cite{wang2020niti,zhou2021octo,wu2016binarized,lin2015neural,rastegari2016xnor, zhou2018adaptive,lin2017towards,courbariaux2015binaryconnect,jacob2018quantization,zhou2016dorefa} and summarize how they manipulate data in Table~\ref{table-low-precision-algorithms}.
We extract the basic concepts from those algorithms and identify 4 key elements that jointly define a mixed-precision training algorithm.
\begin{itemize}[leftmargin=0pt,itemindent=10pt,topsep=0pt]
	\item \textit{Translation from FP32 operators to mixed-precision operators.}
	To support an end-to-end training with mixed precision, a normal FP32 operator often needs to be translated into a combination of new operators that operate on data with different kinds of precision.
	Such translation is elaborately designed by algorithm developers to ensure good accuracy.
	Note that \sys has implemented lots of mixed-precision operators internally.
	\item \textit{Backpropagation rules}
	are used to illustrate how to calculate operators' weight and error gradients.
	For instance, the NITI algorithm uses INT8 deconvolution to calculate the FP32 convolution error gradients.
	\item \textit{Weight information} 
	includes weight type, weight initializing methods, and weight update algorithms.
	\item \textit{Optimizer information}
	includes the loss function (e.g., cross entropy) and the optimizer, like SGD and ADAM.
\end{itemize}

Table~\ref{tab-abstraction} gives an example of how to use the above abstraction to specify the NITI algorithm.
For example, an FP32 convolution needs to be translated into:
(1) an INT8-based convolution;
(2) a Max operation to obtain the scale factor;
(3) a Shift operation to convert INT32 value to INT8 type.
Such a configuration will be an input of \sys along with the model to be trained.
Currently, \sys has already incorporated many built-in mixed-precision training algorithms, i.e., 5 out of 7 in Table~\ref{table-low-precision-algorithms}.
The other two are currently not supported yet due to the lack of support for the certain low-precision operator such as FP8-based convolution, and will be considered in future work.

%

\begin{table}[t]
	\smaller
	\begin{tabular}{c|ccc}
		\hline
		\multirow{4}{*}{Latency compare} & Op  & CPU  & DSP \\ \cline{2-4}
		& Transpose & 3 ms  & 25 ms \\ \cline{2-4}
		& WeightRotate  & 4 ms &   20 ms   \\ \cline{2-4}
		& Slice   &  4 ms & 17 ms  \\  \hline
		DSP unsuppoted ops &  \multicolumn{3}{c}{Normalization, Quantization, Round, Sqrt, etc.} \\ \hline
	\end{tabular}
	\caption{Operators that do not fit to DSP.}
	\vspace{-0.7cm}
	\label{table-op-cpu-dsp}
\end{table}
\subsection{CPU-DSP Co-Scheduling} \label{sec:parallel}
\noindent \textbf{DSP-unfriendly operators}
Though the DSP is adequate to INT8 vector arithmetic operations, there are still some irregular memory access operations or float operations that are unsuitable for DSP to run, referred to as ``\textit{DSP-unfriendly operators}'' in this work.
As shown in Table~\ref{table-op-cpu-dsp}, some operators' latency on DSP is more than 8$\times$ slower than that on CPU, and some FP32-only operators, such as Normalization and Quantization, are lack of architecture-level support on DSP \cite{wang2020niti,yang2020training}.
Therefore, they need to be executed on the CPU.

To partition the model across DSP and CPU, an intuitive approach is to merge the adjacent DSP-unfriendly operators into subgraphs and place them on the CPU.
However, since the CPU-DSP context switching incurs high overhead mainly due to the data copy between their own memory space (i.e., around 25ms on XiaoMI 10), this approach could lead to non-optimal performance.
Conceptually, some DSP-friendly operators shall be placed on the CPU as well to reduce the CPU-DSP context switch frequency.
Therefore, we need a context-switching-aware scheduling strategy that wisely maps the operators to CPU and DSP.

\noindent \textbf{Operator-to-hardware scheduling}
To solve the scheduling problem, \sys first uses topological sort to get an execution order for all operators and profiles to obtain each operator's latency on CPU and DSP.
Then, it uses dynamic programming algorithm to find the approximately optimal scheduling solution while obeying the execution order.
\begin{align}
	T[i+1,CPU] = min\left\{\begin{matrix}
		T[i, CPU] + L_{i+1}^{CPU}\\
		T[i,DSP] + L_{i+1}^{CPU} +L_{switch}
	\end{matrix}\right.
	\label{eq-cpu}
\end{align}
\begin{align}
	T[i+1,DSP] & = min\left\{\begin{matrix}
		T[i, CPU] + L_{i+1}^{DSP} + L_{switch}\\
		T[i,DSP] +L_{i+1}^{DSP}
	\end{matrix}\right.
	\label{eq-dsp}
\end{align}
$T[i+1, CPU]$ is the lowest latency of finishing $Op_{1}.....Op_{i+1}$ if $Op_{i+1}$ running on CPU.
$L_{i+1}^{CPU}$ means the latency of $Op_{i+1}$ running on CPU and $L_{switch}$ represents the latency of context switching. 
The init state is set to $T[1, CPU] = L_1^{CPU}$ and  $T[1, DSP]  = L_1^{DSP}$.
When both $Op_{i+1}$ and $Op_{i}$ running on CPU, there is \textbf{no context switching}.
Otherwise, context switching overhead is added.
$T[i+1, CPU]$ is the minimum value of the above two.
Similarly, the latency of finishing $Op_{i+1}$ running on DSP also has two circumstances as shown in Eq~\ref{eq-dsp}.
The objective $T_{model}$ can be formulated as 
\begin{align}
	T_{model}	= min \{ T[N,CPU], T[N, DSP]\}. 
\end{align}
N is the number of operators.
Based on the recursion formula and the objective, we can find the optimal scheduling plan.
Note that the subgraphs can run on CPU and DSP in parallel, as long as their data dependency is satisfied.
\subsection{Self-Adaptive Rescaling}

\lstset{ %
	basicstyle=\footnotesize,           
	float=t,
	columns=flexible,
	numbers=left,                   
	numberstyle=\tiny\color{gray},  
	stepnumber=1,                   
	numbersep=3pt,                  
	backgroundcolor=\color{white},      
	showspaces=false,               
	showstringspaces=false,         
	showtabs=false,                 
	frame=single,                   
	rulecolor=\color{black},        
	tabsize=2,                      
	language=c, 
	captionpos=b,                   
	breaklines=true,                
	breakatwhitespace=false,        
	title=\lstname,                   
	keywordstyle=\bfseries\color{blue},          
	commentstyle=\footnotesize\it\color[RGB]{0,96,96},       
	stringstyle=\color{mauve},         
	escapeinside={\%*}{*)},            
	morekeywords={Tensor},               
	caption={Key C code snippet of dynamic rescaling},
	label={code:dynamic-rescaling-c}
}
\noindent\begin{minipage}{.28\textwidth}
\begin{lstlisting}
	int scale = 0;
	/*Calculate INT32 temporal results*/
	for(int i = 0; i < length; i++ ) {
		Tensor x = input[i];
		Tensor w = weight[i];
		// CONV or matrix multiply
		Tensor temp_result = x * w; 
		// count leading zero
		Tensor clz = clz(temp_result); 
		int tscale = 32 - max(clz) - 7;
		scale = scale > tscale? scale: tscale;
		temp_output[i] = temp_result; 
	}
	/*  Cast the INT32 to INT8 values */
	for(int i = 0; i < length; i++ ) {
		Tensor temp = temp_output[i];
		// Downscale
		Tensor int8_result = temp / scale;
		result[i]  = int8_result;
	}
\end{lstlisting}
\end{minipage}\hfill
\lstset{ %
	basicstyle=\footnotesize,           
	float=t,
	columns=flexible,
	numbers=left,                   
	numberstyle=\tiny\color{gray},  
	stepnumber=1,                   
	numbersep=3pt,                  
	backgroundcolor=\color{white},      
	showspaces=false,               
	showstringspaces=false,         
	showtabs=false,                 
	frame=single,                   
	rulecolor=\color{black},        
	tabsize=2,                      
	language={[x86masm]Assembler}, 
	captionpos=b,                   
	breaklines=true,                
	breakatwhitespace=false,        
	title=\lstname,                   
	keywordstyle=\bfseries\color{blue},          
	commentstyle=\it\color[RGB]{0,96,96},,       
	stringstyle=\color{mauve},         
	escapeinside={\%*}{*)},            
	morekeywords={loop0,loop1,vmem,vrmpy,vclz,vmax,mux,vmpye},               
	caption={Asm code version},
	label={code:dynamic-rescaling}
}
\begin{minipage}{.17\textwidth}
\begin{lstlisting}
	scale = 0
	
loop0:
	v0 = vmem ptr_i
	v1 = vmem ptr_w
	
	v2 = vrmpy v0, v1
	
	v3 = vclz v2
	tscale =  vmax v3
	scale = mux scale > tscale, scale,tscale
	vmem ptr_t, v2
end loop0
loop1:
	v0 = vmem ptr_t
	
	v3 = vmpye v0, scale
	vmem ptr_v, v3
end loop1
\end{lstlisting}
\end{minipage}
\label{sec:rescaling}

\noindent \textbf{Static scaling vs. dynamic rescaling}
Once a quantized model is deployed for inference, the scale factor (i.e., $S_a^{(l)}$ and $S_w^{(l)}$ in Figure~\ref{fig-int8-training}) per layer is a static value.
Therefore, the data flow is simple as it just needs to multiply the matrix after loading the input and weight, and store the scaled result.
During training, however, the scale factor also needs to be dynamically adapted just like the trainable weights.
An unreasonable scale factor can noticeably drop the model accuracy and the optimal scale factor cannot be known in prior to training completeness.

\noindent \textbf{Mismatch of dynamic rescaling and DSP}
Such dynamic scaling runs slow on DSP for its excessive memory access.
For each batch of training, we have to store the temporal outputs and reload them after obtaining the scale factor to downscale the temporal outputs to final INT8 results, as shown in Listing~\ref{code:dynamic-rescaling-c} and \ref{code:dynamic-rescaling}.
Note that each layer with trainable weights collocates with a scale factor.
Therefore, there can be at most hundreds of dynamic scale factors in a typical DNN model.
As we have measured, dynamic rescaling will at least add 2$\times$ latency compared with a static one.

\begin{figure}[t]
	\centering
	\subfigure[\textbf{Layer's scale factor }] {\includegraphics[width=0.45\textwidth]{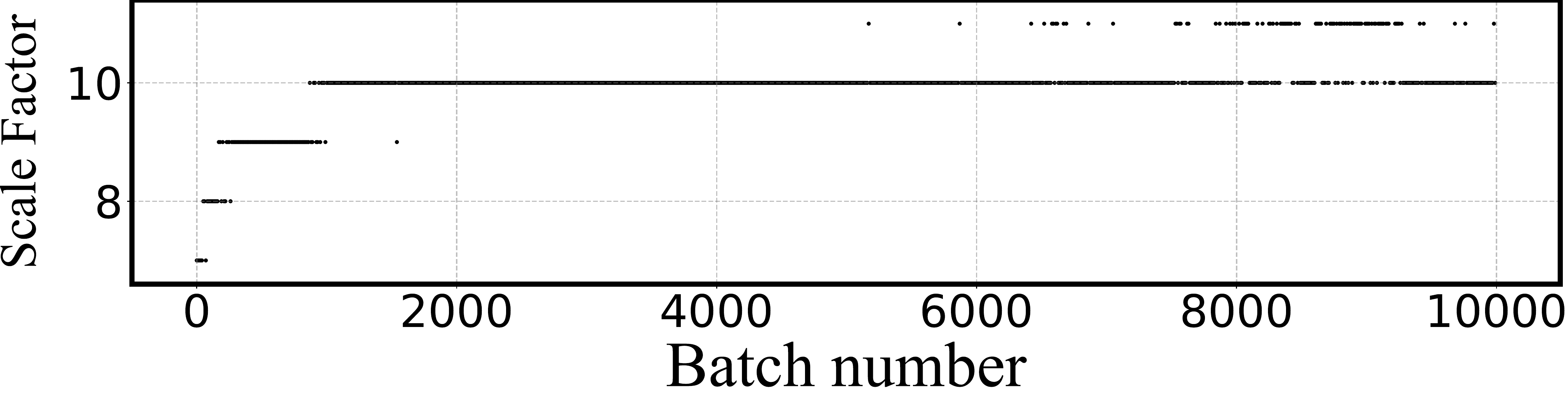} }  \\
\subfigure[\textbf{Layer's scale factor changing interval}] {\includegraphics[width=0.45\textwidth]{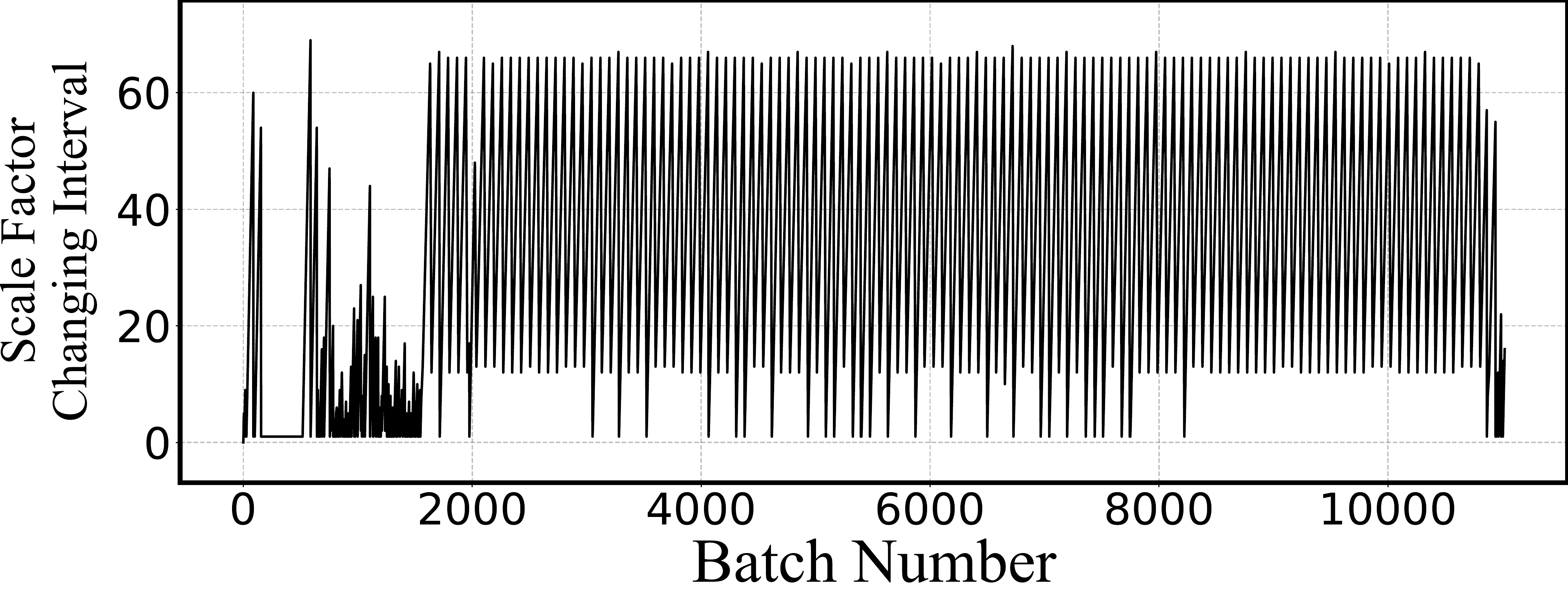} }
\vspace{-0.5cm}
	\caption{The scale factor and its changing interval of the first CONV layer in training  VGG11 model (batch size = 64) on CIFAR-10 dataset. }
	\label{fig-shift}
	\vspace{-0.5cm}
\end{figure}
\noindent \textbf{Insights \& Opportunities}
Fortunately, we observed two useful patterns of how scale factors changes during training.
Figure~\ref{fig-shift} illustrates (a) the concrete value of a scale factor of one layer and (b) its changing frequency in training the VGG11 model on CIFAR-10 dataset.
We find that, after the initialization period:
(1) The scale factor jumps between 2 alternative values, i.e., 10 and 11. Using either of them does not affect the model accuracy.
(2) The actual changing frequency of the scale factor is low, e.g., per 10--60 batches.
Those two phenomenons are common in training because when the model approaches to convergence, the gradients decrease and the scale factor is less likely to be changed.

Based on the above observations, we propose \texttt{self-adaptive rescaling} technique to mitigate the overhead of rescaling.
Its key idea is to periodically enable the rescaling instead of per batch.
The rescaling frequency is adaptively configured based on the observed history of how frequent the scale factors actually change after training, e.g., last $K$ batches with rescaling enabled.
We heuristically set the policy in mapping the observed frequency of changed scale factor $f$ to the periodic frequency in calculating the new scale factor $f/2$.
Luckily, we find such a policy works well for different kinds of models and datasets.

\begin{table}[t]
	\small
	\begin{tabular}{c|rrrrrr}
		\hline
		\multirow{2}{*}{\textbf{Input Size}}  & \multicolumn{6}{c}{\textbf{latencies of different batch sizes (ms)}} \\ \cline{2-7}
		& \textbf{2} & \textbf{4} & \textbf{8} & \textbf{16} &\textbf{32} &\textbf{64}\\ \hline
		8$\times$8 &  0.63 & 0.63 & 0.85 & 1.03 & 1.27 & 1.84 \\ \hline
		16$\times$16 & 0.84 & 0.89 & 4.23 & 3.98 &  \cellcolor{pink}{4.64} & 12.24 \\ \hline
		32$\times$32 & 1.69 & \cellcolor{pink}{2.50} &  \cellcolor{pink}{59.11} &   \cellcolor{pink}{62.35} &  \cellcolor{pink}{68.13} & 152.89 \\ \hline
	\end{tabular}
	\caption{The latency of a convolution layer on DSP with different batch/input sizes (input channel = 64, output channel  = 64). }
	\label{table-batch-size-latency}
	\vspace{-0.7cm}
\end{table}
\subsection{Batch Splitting} \label{sec:splitting}
\noindent \textbf{Exhausted data cache}
A well-known factor towards fast on-device inference is to fully utilize the processor cache~\cite{wang2021asymo}.
For inference tasks, using a large batch is very likely to improve the CPU utilization and, therefore, the processing throughput~\cite{smith2017don,goyal2017accurate}.
However, for DSP-based training tasks, we observe a huge performance decline on the large operator and batch sizes, especially for weight gradients calculation.
Table~\ref{table-batch-size-latency} illustrates this phenomenon with a convolution operator on XiaoMI 10 (Snapdragon 865 SoC).
When the input data size is 32$\times$32, the latency of batch size = 32 is 27.3$\times$ more than that of batch size = 4, which means 8$\times$ theoretical workload incurs 27.3$\times$ delay
\footnote{When batch size > 4, the actual batch size will be padded to a multiple of 32 as required by Hexagon NN, so the latency of batch size = 8 or 16 will be similar to that of batch size = 32.}.

The performance drop comes from the exhausted DSP cache.
There are two reasons for the disparate behavior on training and inference tasks.
First, training tasks often require a large batch size, which leads to numerous amounts of memory access for intermediate data.
Second, DSP cache is often smaller than CPU cache, e.g., only half for L2 cache on Snapdragon 865 (1 MB vs. 2 MB).


To reach a high cache hit ratio, we need to partition the intra-operator workloads.
We choose to split the operation at the batch dimension, i.e., the first dimension of input data, as it is simpler to implement and causes no redundant computations.
We refer an operator to behave ``abnormal'' if its latency-to-workload (in FLOPs) ratio is noticeably higher than the same configuration but with a smaller batch size.
Through offline profiling, like we have done in Table~\ref{table-batch-size-latency}, \sys can identify all abnormal operators in a model and split them into normal ones.

To ensure efficient locality, \sys splits an abnormal batch into multiple micro-batches and executes them individually.
The final weight gradients are the accumulation of their output. 
The accumulation formula for FP32 operations is $W^g = \sum^n_{i=1} W^g_{batch_i}$.
When comes to INT8, the formula will be changed into $W^g*S^g = \sum^n_{i=1}W^g_{batch_i}*S^g_{batch_i}$.
This formula can be finally transformed to 
\vspace{-0.2cm}
\begin{align}
	W^g = \sum^n_{i=1} W^g_{batch_i}*S^g_{batch_i}/S^g
	\label{eq-split}
\end{align}
\vspace{-0.3cm}

According to Eq~\ref{eq-split}, if $S^g_{batch_*}/S^g = 1$, we can avoid the FP32 add operation.
Refer to non-split gradient algorithm, we can know that $S^g =  max \{  S^g_{batch_i}, i  \in split\_num   \} $.
Therefore, all the temporary micro-batch results should rescale from $S^g_{batch_i}$ to $S^g$.
Our experiments show that when splitting the batch,  in most cases $S^g_{batch_i}$  is the same as $S^g$, so rescaling will not compromise the benefits from batch splitting.

\subsection{Compute Subgraph Reuse} \label{sec:reuse}
\noindent \textbf{Costly compute graph preparation}
Our experiments also show that preparing the compute graph takes a considerable amount of time, e.g., 304ms on TFLite and 212ms on MNN for the VGG16 \cite{simonyan2014very} model, respectively.
The preparation includes the following steps.
First, the training engine needs to build the DSP compute subgraph, which consists of operators with inputs, outputs, and parameters.
To ensure that the subgraph can execute correctly, the engine also maintains a reference relationship graph between operators.
Before invoking the DSP training, the memory space needs to be allocated on DSP as well.
The current on-device training engines always prepare a new compute subgraph for each batch of training. The dynamic graphs are rather easy for developers to debug and develop, however, unfortunately they incur high overhead on resource-constrained devices, which are not suitable for on-device training. 

Since the models are rarely modified during on-device training, we propose to reuse the DSP compute subgraph to eliminate its preparation overhead.
However, directly reusing the subgraph can easily exceed the DSP memory budget since substantial memory regions cannot be released.
To this end, \sys seeks to minimize the memory allocation/deallocation operations under the memory constraint -- a common memory management problem in the operating system~\cite{zhao2005dynamic,pyka2007operating,tiwari2010mmt,beyler2007performance}.
An opportunity is that, unlike OS, \sys's memory allocation/deallocation for subgraph reuse always follows DNN's execution order, so that the memory region most recently used (MRU) has the longest reuse distance.
Therefore, the key idea is to release the MRU memory regions which best fit memory needs.
\begin{itemize}[leftmargin=0pt,itemindent=10pt,topsep=0pt]
	\item At the preparing stage, \sys profiles all memory regions that compute subgraphs use as shown in Figure~\ref{fig-workflow}.
	Since the number of compute subgraphs is rather small ($<$100), we can exhaustively explore all circumstances to find all possible solutions satisfying different requiring memory sizes.
	\item When the system is about to exceed the memory budget, \sys releases the MRU memory regions identified and marked at the preparing stage, and allocates space for newly coming subgraphs.
\end{itemize}

	\section{Implementation and Evaluation}\label{sec:eval}
We have fully implemented \sys with 15k LoC in C/C++ and 800 LoC of assembly language in total.
The prototype is a standalone framework supporting models exported from MNN \cite{mnn}, TFLite \cite{tflite}, and Pytorch Mobile \cite{pytorch}.
\sys leverages Hexagon NN \cite{hexagon-nn} as the DSP backend, which is the only  open-source library supporting inference on DSP developed by Qualcomm.
While \sys has supported many different mixed-precision training algorithms as shown in Table~\ref{tab-abstraction}, in our experiments, we use NITI~\cite{wang2020niti} as the default one, because it extensively uses INT8 operations that are quite suitable for DSP.
Since Hexagon DSP architecture is constantly changing, we mainly optimize our implementation for V66 architecture through assembly code.
The prototype reuses the FP32-based operators running on the CPU from \texttt{MNN}.
\subsection{Experimental Methodology}

\begin{table}[t]
	\smaller[2]
			\setlength{\tabcolsep}{1mm}{
			\begin{tabular}{c|c|c|c}
				\hline
				 \textbf{Devices} &  \textbf{CPU} & \textbf{GPU} & \textbf{DSP} \\  \hline
				 \multirow{3}{*}{  \makecell[c]{XiaoMI 11 Pro \\ Snapdragon  888}}
				&   2.84GHz Cortex-X1& \multirow{3}{*}{  \makecell[c]{ Adreno 660 GPU \\ 700MHz}} & \multirow{3}{*}{  \makecell[c]{ Hexagon 780 DSP \\ 500MHz}}   \\ 
				& 3$\times$ 2.4GHz Cortex A78 &   &   \\  
				& 4$\times$ 1.8GHz Cortex A55 &   & \\[3pt]  \hline
				
				\multirow{3}{*}{  \makecell[c]{XiaoMI 10 \\ Snapdragon  865}}
				&   2.84GHz A77&   \multirow{3}{*}{  \makecell[c]{Adreno 650 GPU \\  587MHz}}& \multirow{3}{*}{  \makecell[c]{Hexagon 698 DSP \\ 500MHz}}   \\ 
				& 3$\times$ 2.4GHz Cortex A77 &   &  \\  
				& 4$\times$ 1.8GHz Cortex A55 &   &  \\[3pt]  \hline
				
				 \makecell[c]{Redmi Note9 Pro \\ Snapdragon  750G}
				&    \makecell[c]{2$\times$ 2.2GHz Cortex A77 \\ 6$\times$ 1.8GHz Cortex A55 } 
				& \makecell[c]{Adreno 619 GPU  \\ 950MHz}   
				&  \makecell[c]{Hexagon 694 DSP   \\ 500MHz}  \\[3pt]  \hline
		\end{tabular}}
	\vspace{2pt}
	\caption{Devices used in the experiments.}
	\label{table-devices}
	\vspace{-0.7cm}
\end{table}
\begin{table}[t]
	\small
	\begin{tabular}{llrr}
		\hline
		\textbf{Model} & \textbf{Input Data} & \textbf{FLOPs} & \textbf{\# of CONVs} \\ \hline
		VGG-11~\cite{simonyan2014very} & CIFAR-10 & 914 M  & 8 \\ \hline
		VGG-16~\cite{simonyan2014very} &  CIFAR-10& 1.35 G & 13 \\ \hline
		VGG-19~\cite{simonyan2014very} & ImageNet & 26.92 G & 16 \\ \hline
		ResNet-34~\cite{he2016deep} & CIFAR-10 & 7.26 G & 36 \\ \hline
		ResNet-18~\cite{he2016deep} & ImageNet & 11.66 G & 20 \\ \hline
		InceptionV3~\cite{szegedy2016rethinking} &  CIFAR-10 & 2.43 G& 16 \\ \hline
	\end{tabular}
\caption{DNN models used in the experiments.}
\label{table-models}
\vspace{-0.4cm}
\end{table}

\begin{figure*}[t]
	\centering
		 \includegraphics[width=0.95\textwidth]{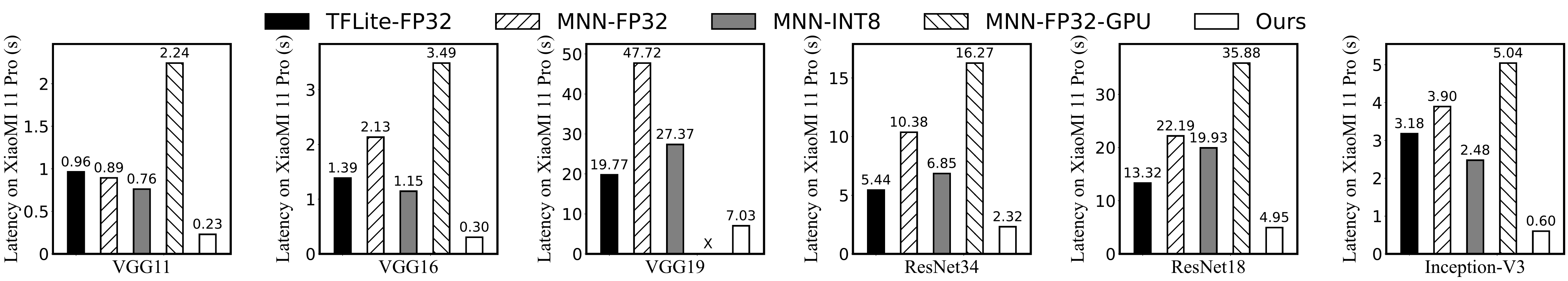} \\  
		\includegraphics[width=0.95\textwidth]{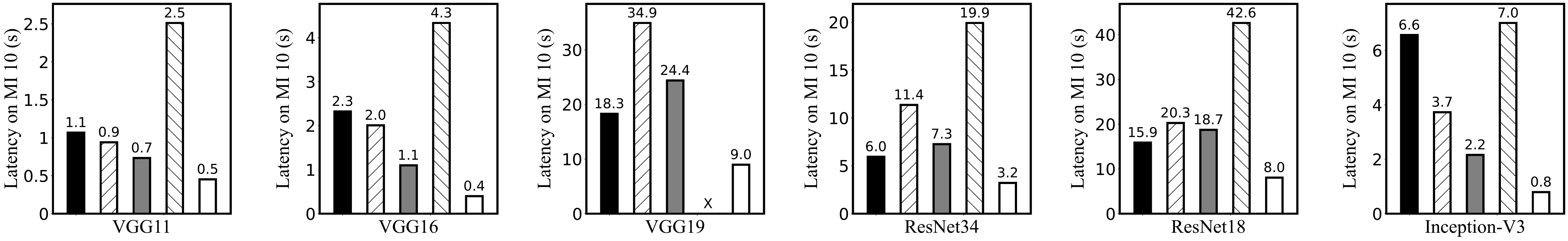}  \\
		\includegraphics[width=0.95\textwidth]{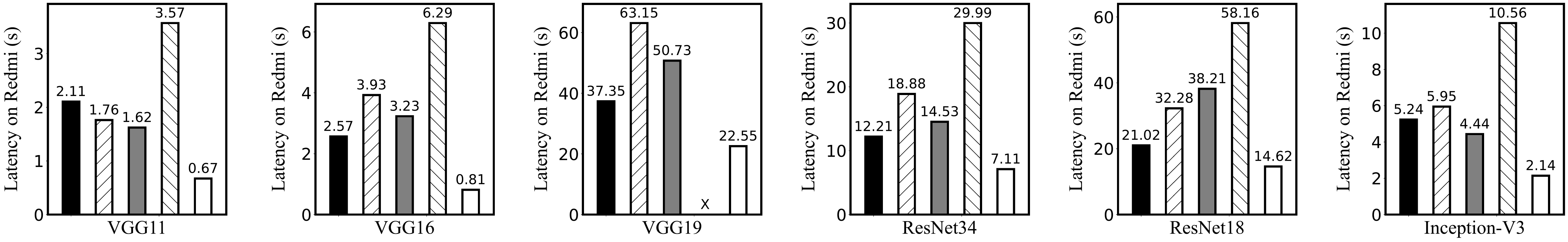} 
	\vspace{-0.4cm}
	\caption{Per-batch training time on different models (batch size = 64) on different devices.}
	\label{fig-per-batch-latency}
	\vspace{-0.5cm}
\end{figure*}

\noindent \textbf{Hardware setup.}
We test the performance of \sys on three smartphones with different Qualcomm SoCs: XiaoMI 11 Pro (Snapdragon 888), XiaoMI 10 (Snapdragon 865), and Redmi Note9 Pro (Snapdragon 750G).
The XiaoMI 11 Pro device is equipped with the latest Hexagon 780 DSP, which is claimed to have huge performance promotions over the old ones.
The hardware details are shown in Table~\ref{table-devices}.
All devices run Android OS 10.
By default, we always run the baselines on 4 BIG CPU cores and \sys's CPU workloads on 2 BIG cores.
The CPU frequency is controlled by OS's dynamic voltage and frequency scaling (DVFS) controller.

\noindent \textbf{Models.}
We test with a range of typical CNN models with various input sizes: VGG11/16/19~\cite{simonyan2014very}, ResNet18/34~\cite{he2016deep}, and InceptionV3~\cite{szegedy2016rethinking}, as listed in Table~\ref{table-models}.
The input data to those models are either CIFAR-10~\cite{krizhevsky2009learning} (input size 32x32) or ImageNet~\cite{krizhevsky2012imagenet} (input size 224x224).

\noindent \textbf{Baselines.}
We mainly compare \sys with two frameworks.
One is MNN~\cite{mnn} that is one of the earliest frameworks that has supported on-device training since late 2019.
Note that \sys also reuses some operator implementation from MNN.
The other one is TFLite~\cite{tflite}, which is the most popular DL framework on smartphones and lately added training support in November, 2021.
Both MNN and TFLite only support training with FP32 format.
To make a more fair comparison, we also extend MNN to train with INT8 format using the same training algorithm as \sys on CPU.
More specifically, we compare \sys with four baselines: 
(1) \texttt{TFLite-FP32}: the traditional FP32-based training method provided by TFLite.
(2) \texttt{MNN-FP32}: the traditional FP32-based training method provided by MNN.
(3) \texttt{MNN-INT8}: the INT8-based training method implemented by us based on MNN.
NEON~\cite{reddy2008neon} is extensively used to optimize the performance of this baseline.
(4) \texttt{MNN-FP32-GPU}: FP32-based training on mobile GPU through OpenCL backend.

\begin{figure*}[t]
	\centering
	
	\includegraphics[width=0.95\textwidth]{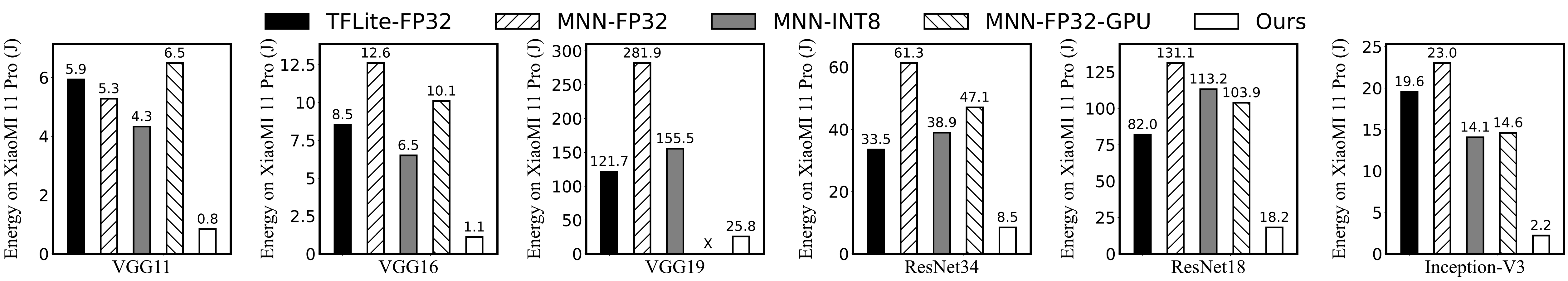} \\  
	\includegraphics[width=0.95\textwidth]{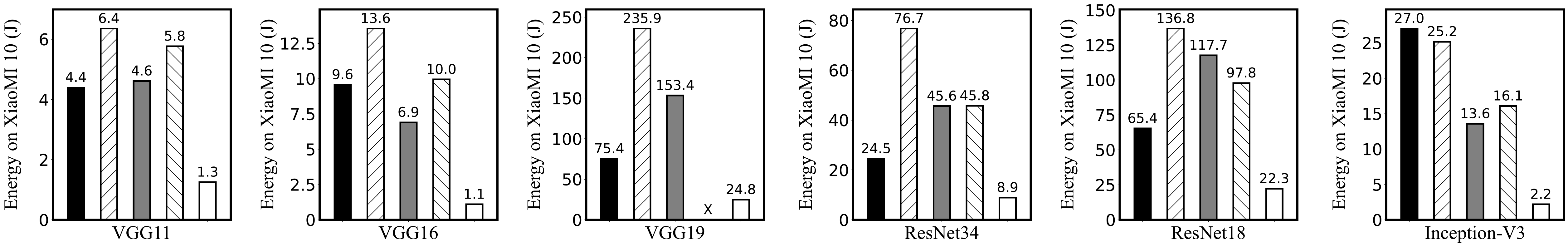}  \\
	\includegraphics[width=0.95\textwidth]{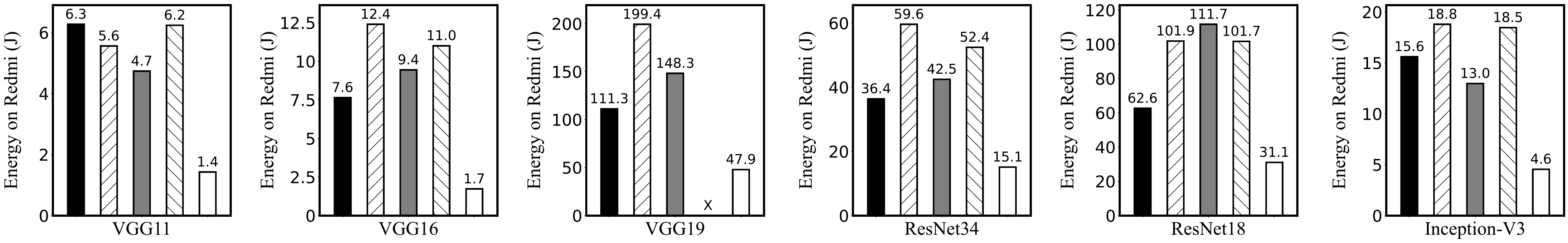} 
	\vspace{-0.3cm}
	\caption{Per-batch energy consumption on different models (batch size = 64) on different devices.}
	\label{fig-per-batch-energy}
	\vspace{-0.5cm}
\end{figure*}

\noindent \textbf{Metrics and configurations.}
We mainly measure the training time and energy consumption during training.
The energy consumption is calculated through Android’s vFS (\texttt{/sys /class/power\_supply}) by profiling every 100ms.
Besides, we also evaluate thermal and CPU frequency  through \texttt{/sys/ class/thermal/thermal\_zone} and \texttt{/sys/devices/system/ cpu/cpufreq} to show our power efficiency from long duration of intensive computation.
All experiments are repeated by 3 times and the average numbers are reported.

\subsection{Per-Batch Performance}

\textbf{Overall performance}
We first comprehensively investigate the per-batch training performance of \sys with batch size 64.
The latency and energy consumption results are illustrated in Figure~\ref{fig-per-batch-latency} and Figure~\ref{fig-per-batch-energy}, respectively.
Our key observation is that \sys \textit{ consistently and remarkably outperforms other baselines on both metrics.}
\begin{itemize}[leftmargin=0pt,itemindent=10pt,topsep=0pt]
	\item \textit{Training latency of \sys vs. FP32 baselines}
	Compared with MNN-FP32 and TFLite-FP32, \sys achieves 2.08-7.1$\times$ and  1.44-8.25$\times$ speedup of per-batch training time, respectively.
	Comparing different devices, we observe that \sys's improvements are relatively less profound on Redmi Note9 Pro than the other two devices.
	This is because Redmi Note9 Pro is equipped with an outdated SoC where the performance gap between CPU and DSP is much smaller than the other two high-end SoCs.
	
	\item \textit{Energy consumption of \sys vs. FP32 baselines}
	As Figure~\ref{fig-per-batch-energy} shows, \sys's improvements on energy consumption are even more profound than training speed.
	Specifically, \sys reduces the energy consumption by 3.21-11.2$\times$ and 2.01-12.5$\times$ compared with MNN-FP32 and TFLite-FP32, respectively.
	Such a huge benefit comes from both the training speedup and the higher power efficiency of DSP.
	
	\item \textit{\sys vs. MNN-FP32-GPU}
	\sys can reduce 3.98-11.63$\times$ latency and  3.46-10.95$\times$ energy consumption compared with MNN-FP32-GPU.
	The reason for such huge improvement is that 
	(1) As far as we know, MNN is the only framework supporting on-device GPU training and is not fully optimized yet.
	The GPU utilization during training is only around 30-50\%.  
	(2) DSP is more power-efficient than GPU.
	Besides, training the VGG19 model on ImageNet with MNN-FP32-GPU encounters out-of-memory failure.
	
	\item \textit{\sys vs. MNN-INT8}
	According to Figure~\ref{fig-per-batch-latency} and Figure~\ref{fig-per-batch-energy}, \sys can reduce up to 4.13$\times$ latency and 6.5$\times$ energy consumption compared to MNN-INT8, respectively.
	Since both of them use the same mixed-precision training algorithm, the improvements come from \sys's ability to fully utilize the DSP hardware.
	Note that DSP HVX vector instruction can at most calculate 128 INT8 arithmetic operations; while CPU NEON can only perform 4 operations.
\end{itemize}


\noindent \textbf{Impacts of batch size}
\begin{figure}[t]
	\centering
	\subfigure[\textbf{VGG16 Latency}]{ \includegraphics[width=0.22\textwidth]{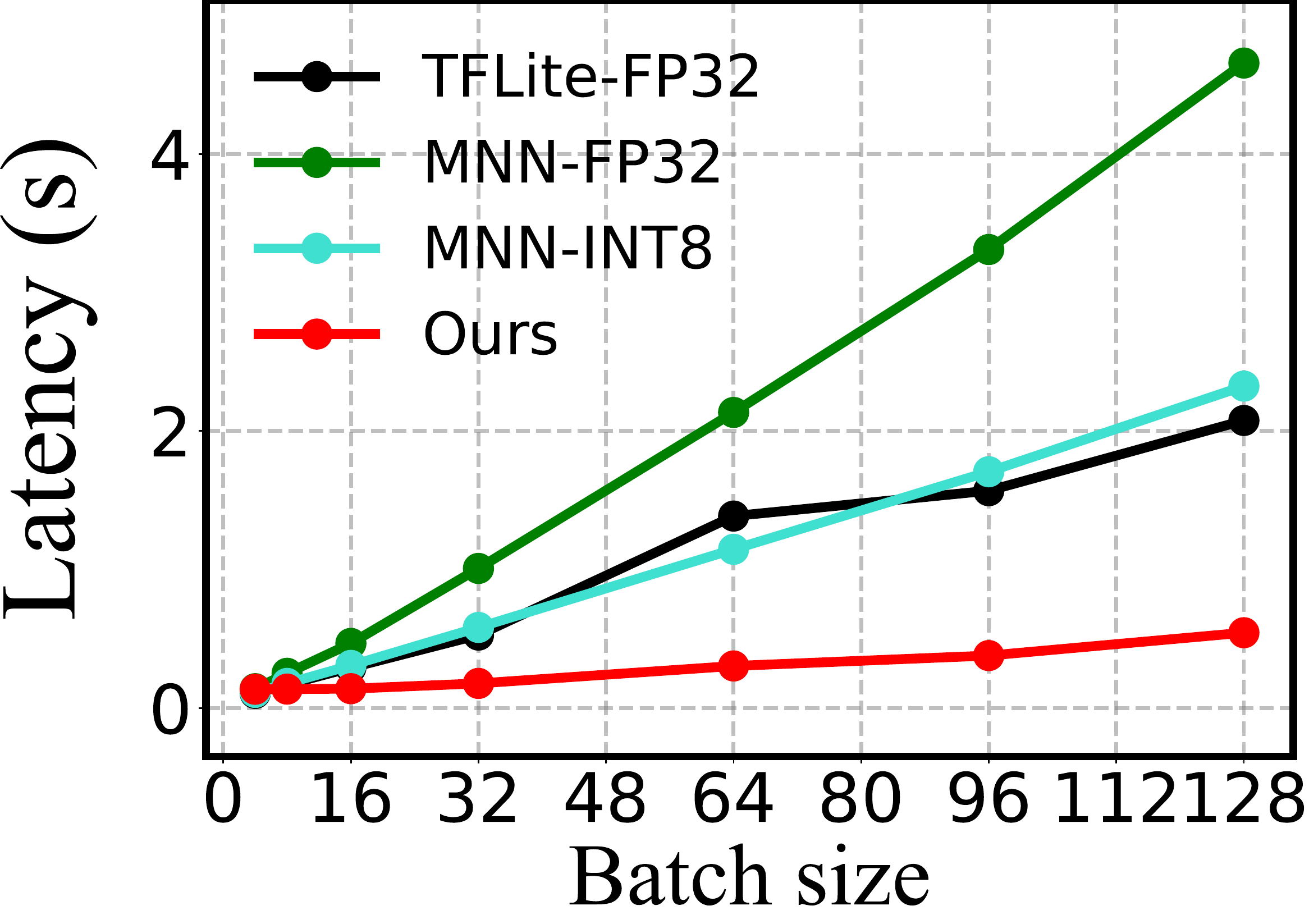}  } 
	\subfigure[\textbf{VGG16 Energy}]{ \includegraphics[width=0.22\textwidth]{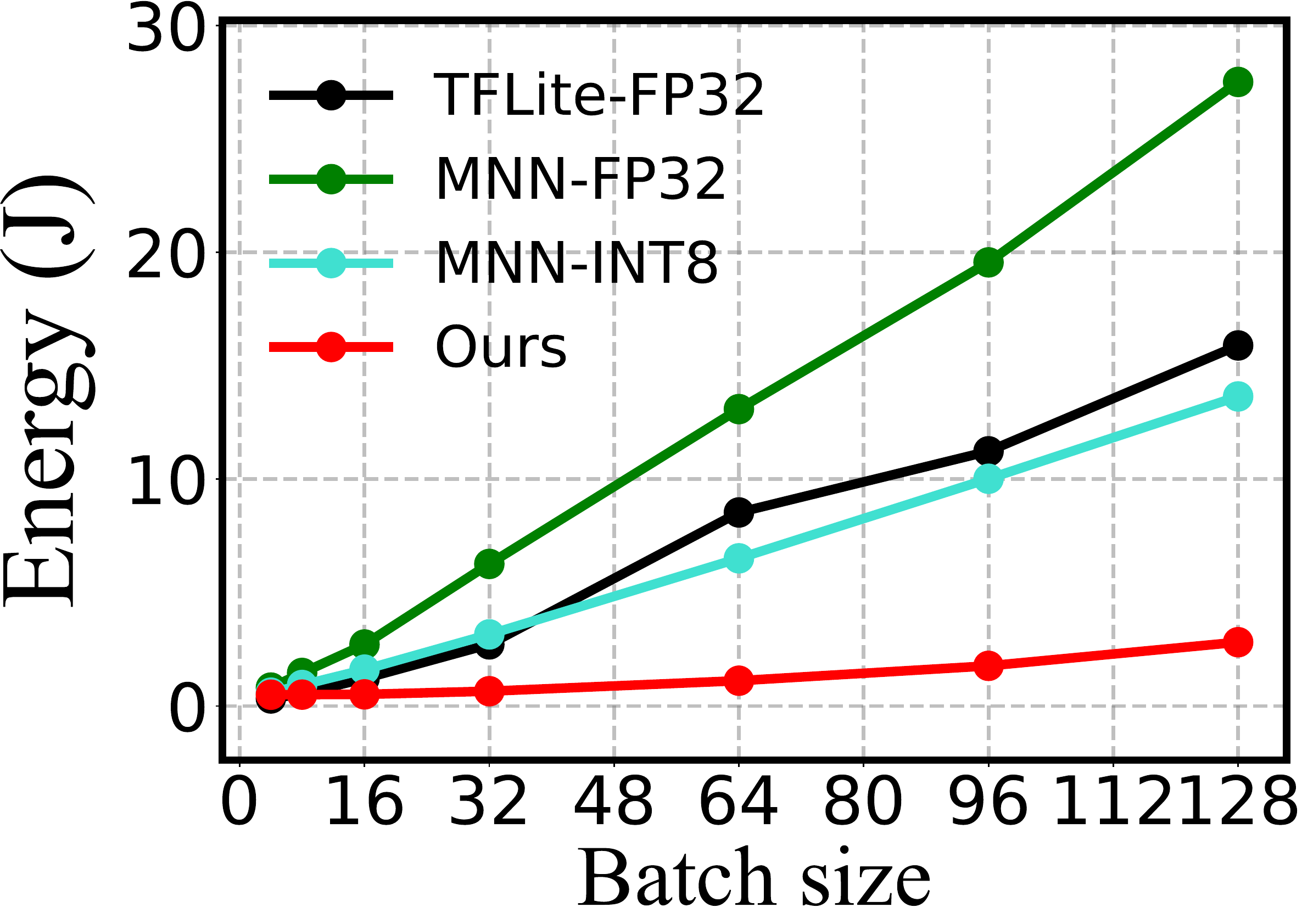}  } 
	\subfigure[\textbf{InceptionV3 Latency}]{ \includegraphics[width=0.22\textwidth]{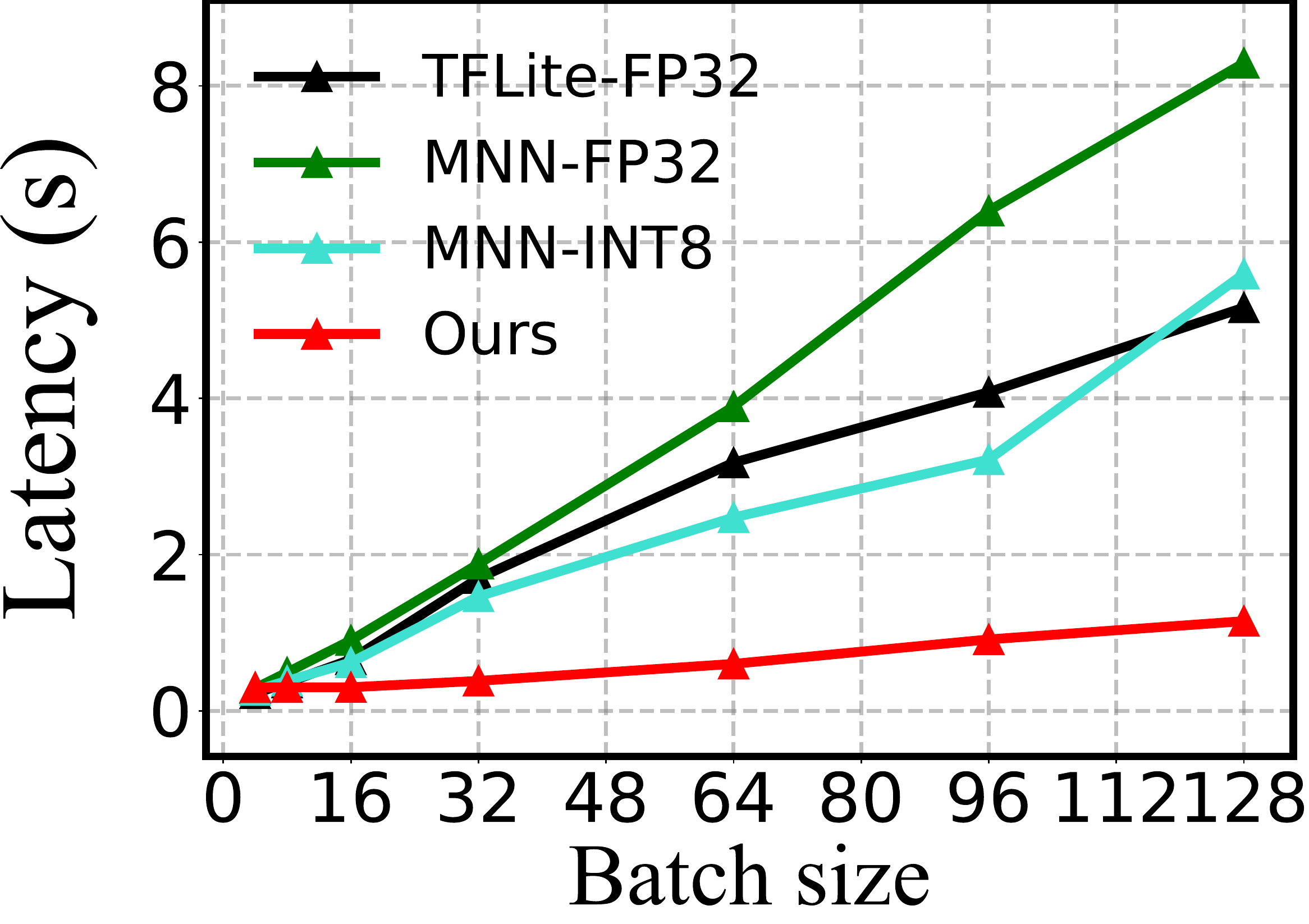}  } 
	\subfigure[\textbf{InceptionV3 Energy}]{ \includegraphics[width=0.22\textwidth]{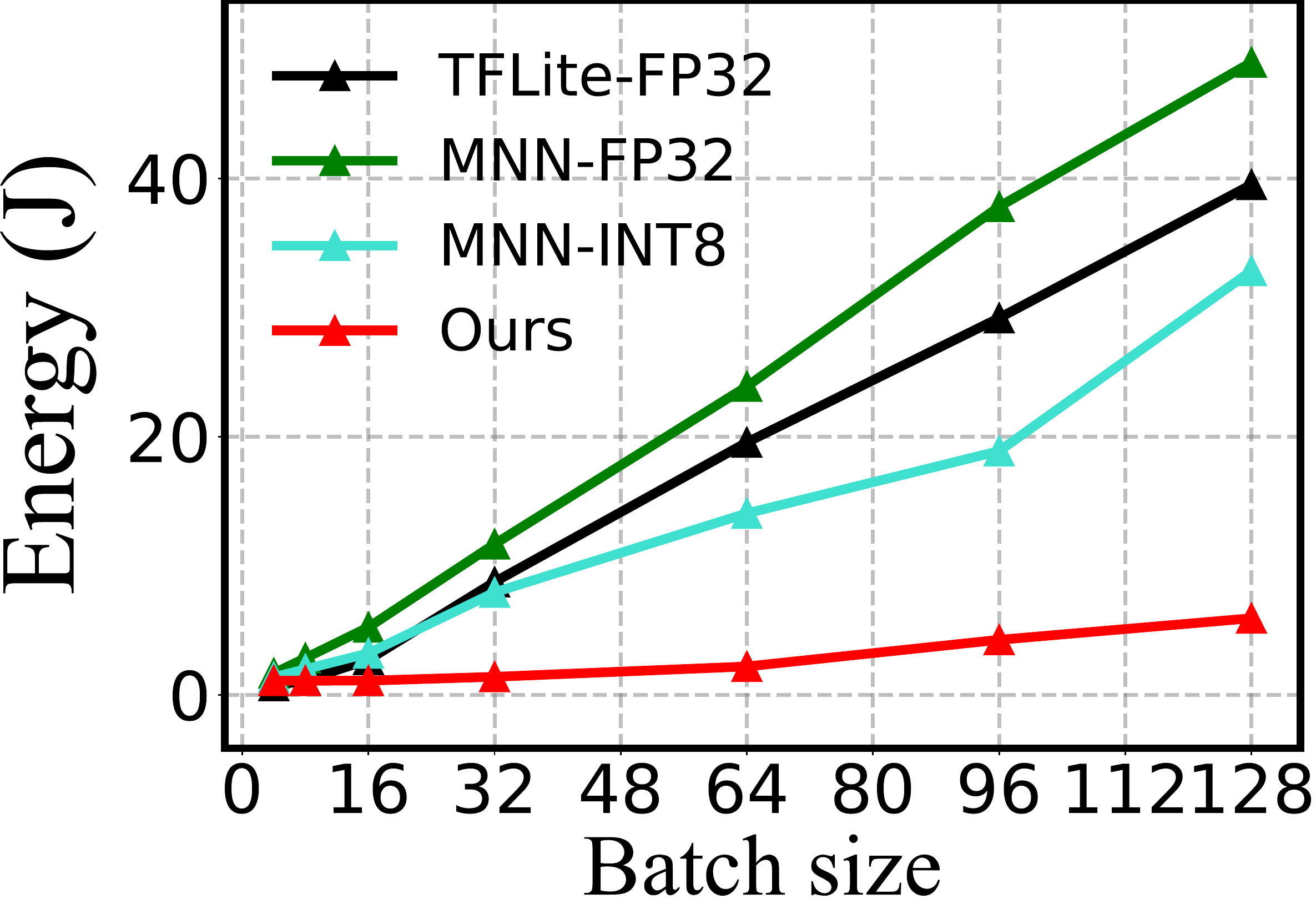}  } 
	\vspace{-0.5cm}
	\caption{Per-batch training time and energy consumption  under various batch sizes.} 
	\label{fig-batchsize}
	\vspace{-20pt}
\end{figure}
We then evaluate the performance of \sys with various batch sizes from 4 to 128 with VGG16 and InceptionV3 models on XiaoMI 11 Pro.
As shown in Figure~\ref{fig-batchsize}, \sys consistently outperforms other baselines on each batch size, e.g., 4.81$\times$ lower latency and 6.90$\times$ lower energy consumption on average.
Besides, the performance gap between \sys and baselines is bigger with batch size increasing.
For instance, \sys reduces up to 8.56$\times$ latency and 12.67$\times$ energy consumption for batch size 128 thanks to the batch splitting technique ($\S$\ref{sec:splitting}).


\noindent\textbf{Impacts of CPU cores}
\begin{table}[t]
	\footnotesize
	
	\begin{center}
	\begin{tabular}{c|c|rr|rr}
		\hline
		\multirow{2}{*}{\textbf{Method}} & \multirow{2}{*}{\textbf{CPU Conf.}} & \multicolumn{2}{c|}{\textbf{Time (s)}} & \multicolumn{2}{c}{\textbf{Energy (J)}} \\\cline{3-6} 
		
		&  & H & L & H & L \\ \hline

		\multirow{6}{*}{\rotatebox{90}{MNN-FP32}} & BIG 1$\times$ & 4.88 & 10.81 &  11.05  & \cellcolor{pink}{6.90}  \\ \cline{3-6} 
		& BIG 2$\times$ & 2.87 & 5.89  &  12.85 &  7.50 \\ \cline{3-6} 
		& BIG 4$\times$ (default) &  \cellcolor{pink}{2.14} & 3.86 &11.90 & 7.10 \\ \cline{3-6} 
		& LITTLE 1$\times$ &25.20  & 42.57 & 13.35 &  21.95 \\ \cline{3-6} 
		& LITTLE 4$\times$ & 10.75 & 17.06 & 14.85  & 9.55 \\ \cline{3-6} 
		& Hybrid 8$\times$ & 4.50  & 7.58 & 25.40  & 12.45 \\ \hline

		\multirow{6}{*}{\rotatebox{90}{MNN-INT8}} & BIG 1$\times$ &  5.59 & 14.75 & 9.85 &  \cellcolor{pink}{4.35}  \\ \cline{3-6} 
		& BIG 2$\times$ & 3.07 & 7.37  & 10.60  & 5.25 \\ \cline{3-6} 
		& BIG 4$\times$ (default) &  \cellcolor{pink}{1.79}  & 3.8 & 7.90 & 4.55 \\ \cline{3-6} 
		& LITTLE 1$\times$ & 40.13 &  81.20 & 12.65  & 25.85 \\ \cline{3-6} 
		& LITTLE 4$\times$ &  10.36 &  20.64 & 10.35  &  5.95 \\ \cline{3-6} 
		& Hybrid 8$\times$ & 2.71  & 4.96 & 12.10  & 6.20 \\ \hline

		\multirow{6}{*}{\rotatebox{90}{Ours}} & BIG 1$\times$ & 0.29 & 0.41  &0.70 &  \cellcolor{pink}{0.65}  \\ \cline{3-6} 
		& BIG 2$\times$ (default) & 0.25  &  0.32 &  0.90 & 0.70 \\ \cline{3-6} 
		& BIG 4$\times$ & \cellcolor{pink}{0.24} &  0.31 & 1.20 & 0.90  \\ \cline{3-6} 
		& LITTLE 1$\times$ & 0.79 & 1.61 & 0.80 &  1.15 \\ \cline{3-6} 
		& LITTLE 4$\times$ &  0.49 & 0.68 &   0.90 & 0.75 \\ \cline{3-6} 
		& Hybrid 8$\times$ &  0.34 & 0.39 & 1.60 & 0.90 \\ \hline

	\end{tabular}
\end{center}
	\caption{The performance impacts from the selection of CPU cores and their frequency (H/L: highest/lowest frequency available).
		Numbers in highlight indicate the best performance or least energy consumption. 
		}
	\vspace{-30pt}
	\label{tab-big-little}
\end{table}
Since most baselines and part of \sys runs on CPUs, it's intuitive to test how the choice of CPU cores affect their performance.
In this experiment we use the VGG16 model with batch size 64.
We vary the CPU core numbers 
and the frequency for each core on XiaoMI 11 Pro.
The results are summarized in Table~\ref{tab-big-little}.

As observed, the choice of CPU cores opens rich tradeoffs for \sys and baselines between the training speed and energy consumption.
Running on 4 BIG CPU cores with the highest frequency enables \sys to train with the fastest speed (0.24s per batch), yet its energy consumption is 1.9$\times$ higher than 1 BIG CPU core with the lowest frequency.
The default configuration for \sys is to use 2 BIG CPU cores with DVFS to make a balance between the two key metrics.
In reality, a developer or the OS might control the CPU cores to harness such a tradeoff.
For instance, on a low-power device, the OS might signal \sys to use only one BIG CPU core for training.
To be noted, \sys still significantly outperforms other baselines with any settings.

\subsection{End-to-End Convergence}

\begin{figure}[t]
	\centering
	\subfigure[\textbf{VGG11-CIFAR10-Local}]{ \includegraphics[width=0.22\textwidth]{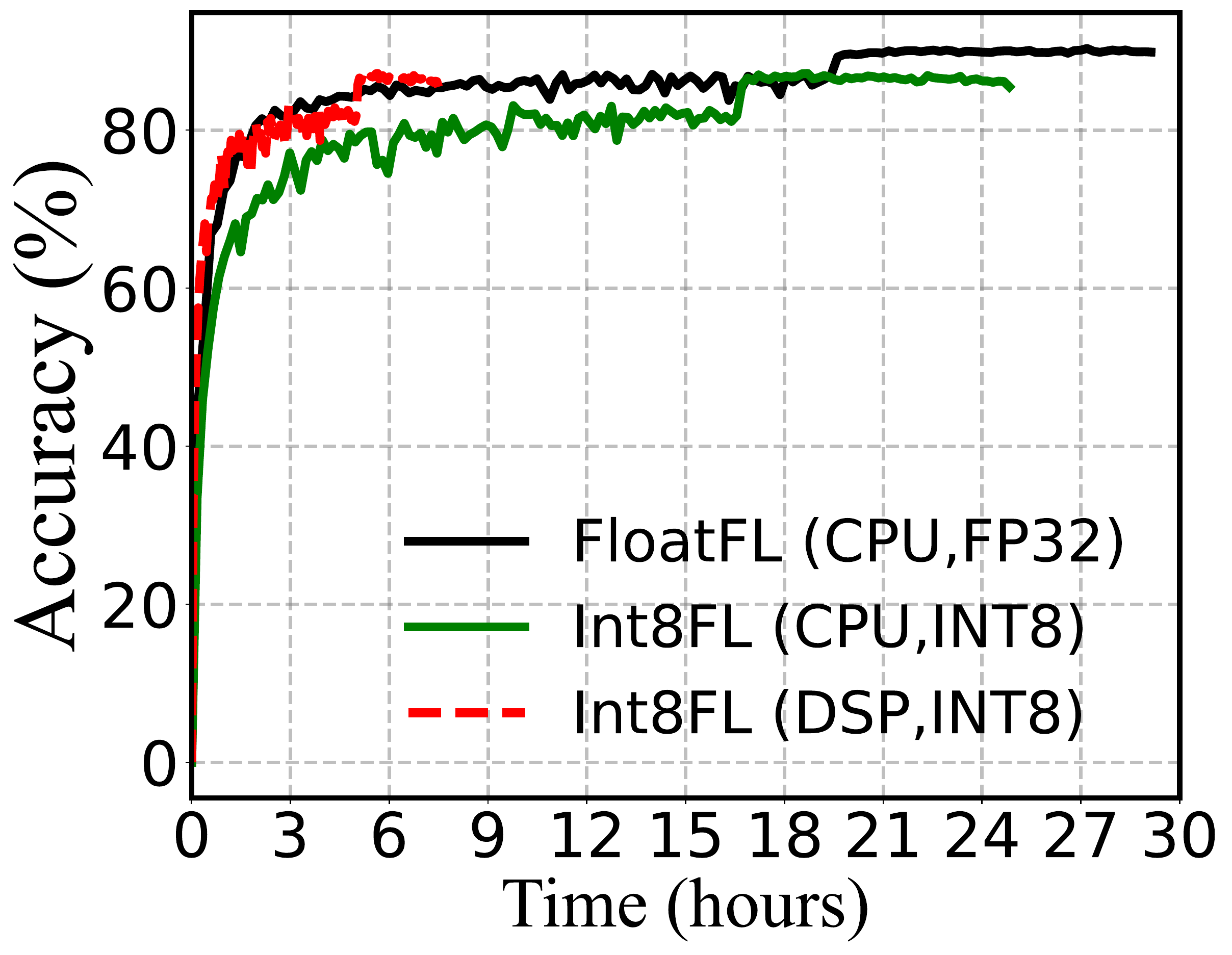} }
	\subfigure[\textbf{ResNet18-CIFAR10-Local}]{ \includegraphics[width=0.22\textwidth]{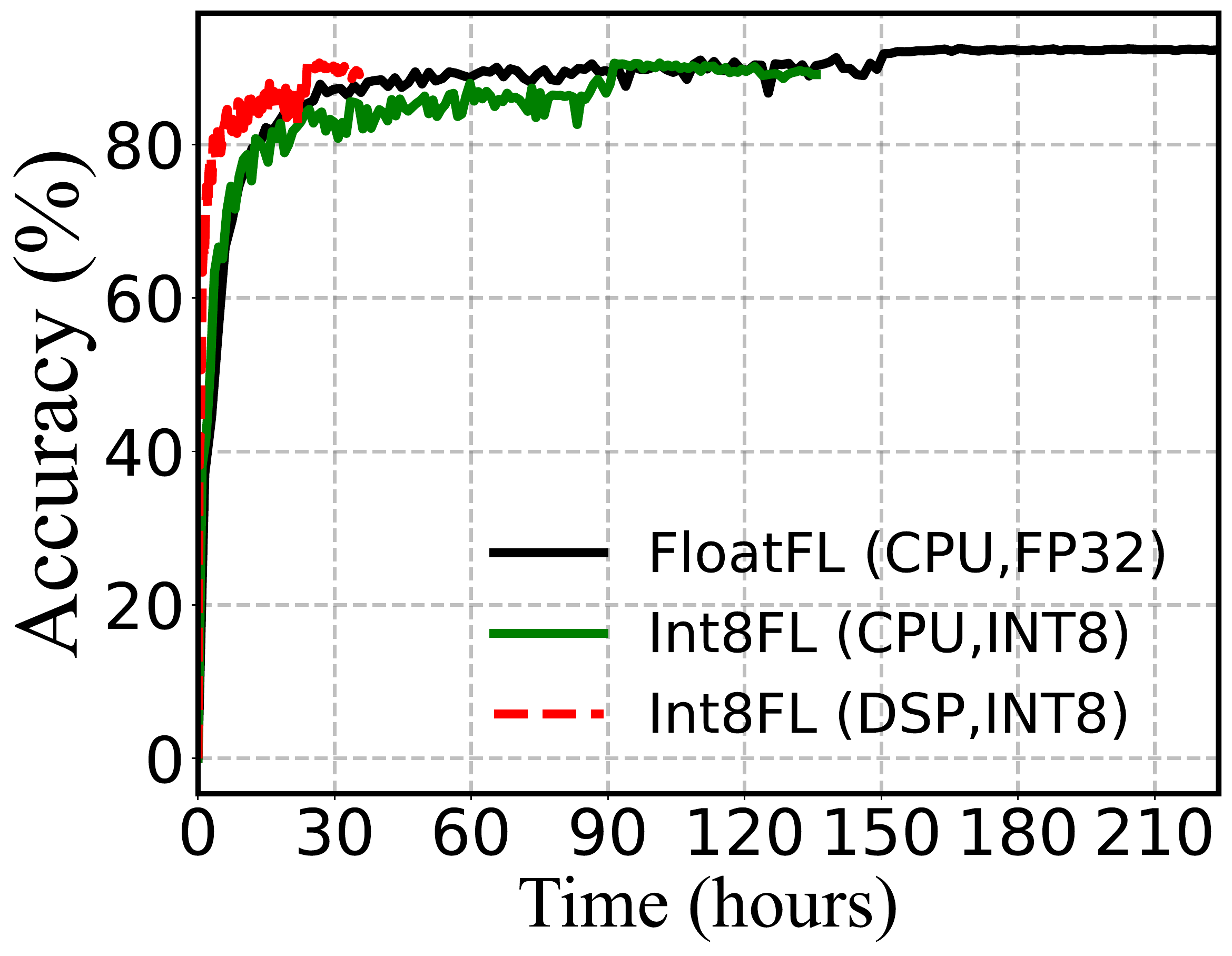} } \\ 
	\subfigure[\textbf{LENET-FEMNIST-Federated}]{ \includegraphics[width=0.22\textwidth]{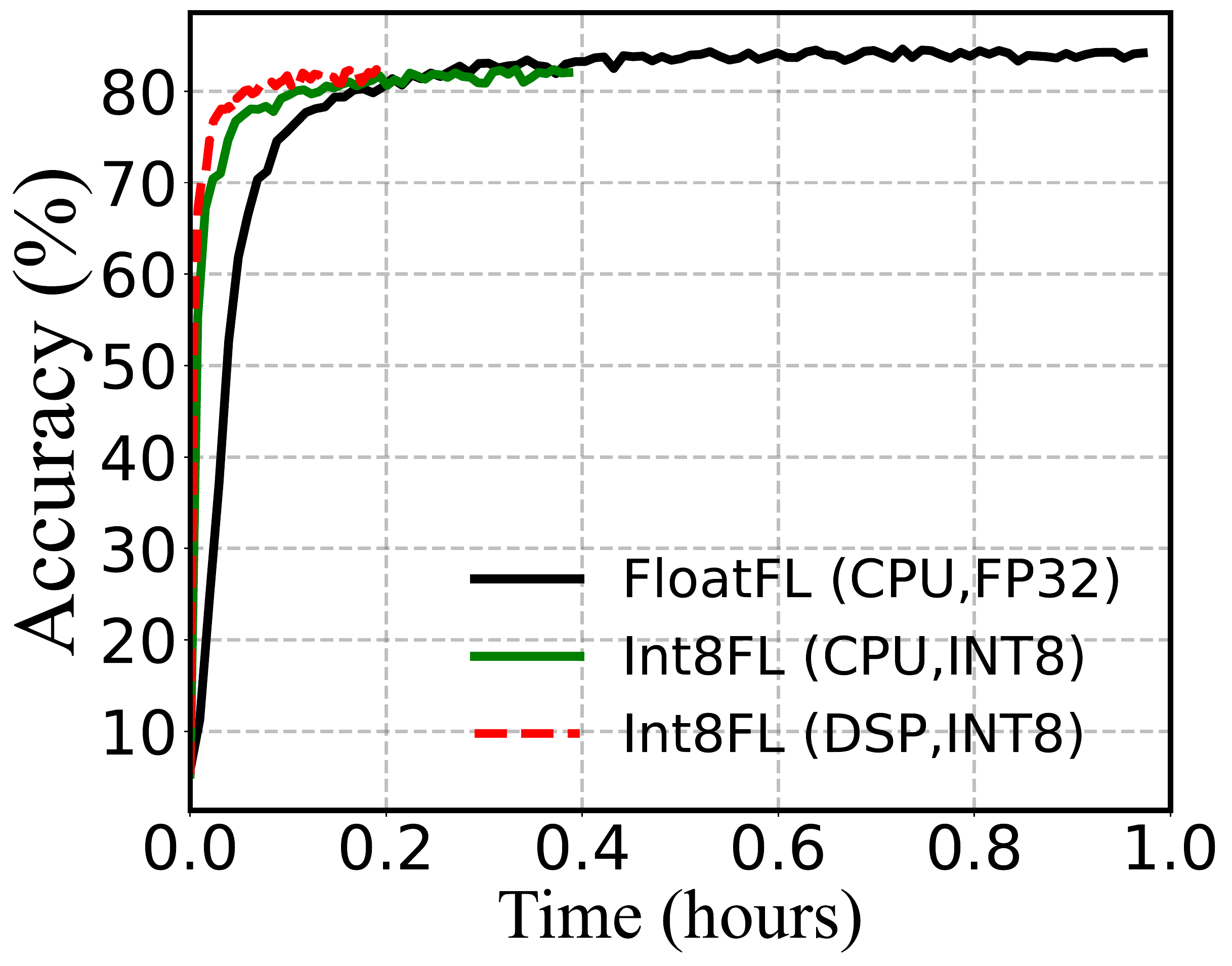}  } 
	\subfigure[\textbf{VGG16-CIFAR100-Federated}]{ \includegraphics[width=0.22\textwidth]{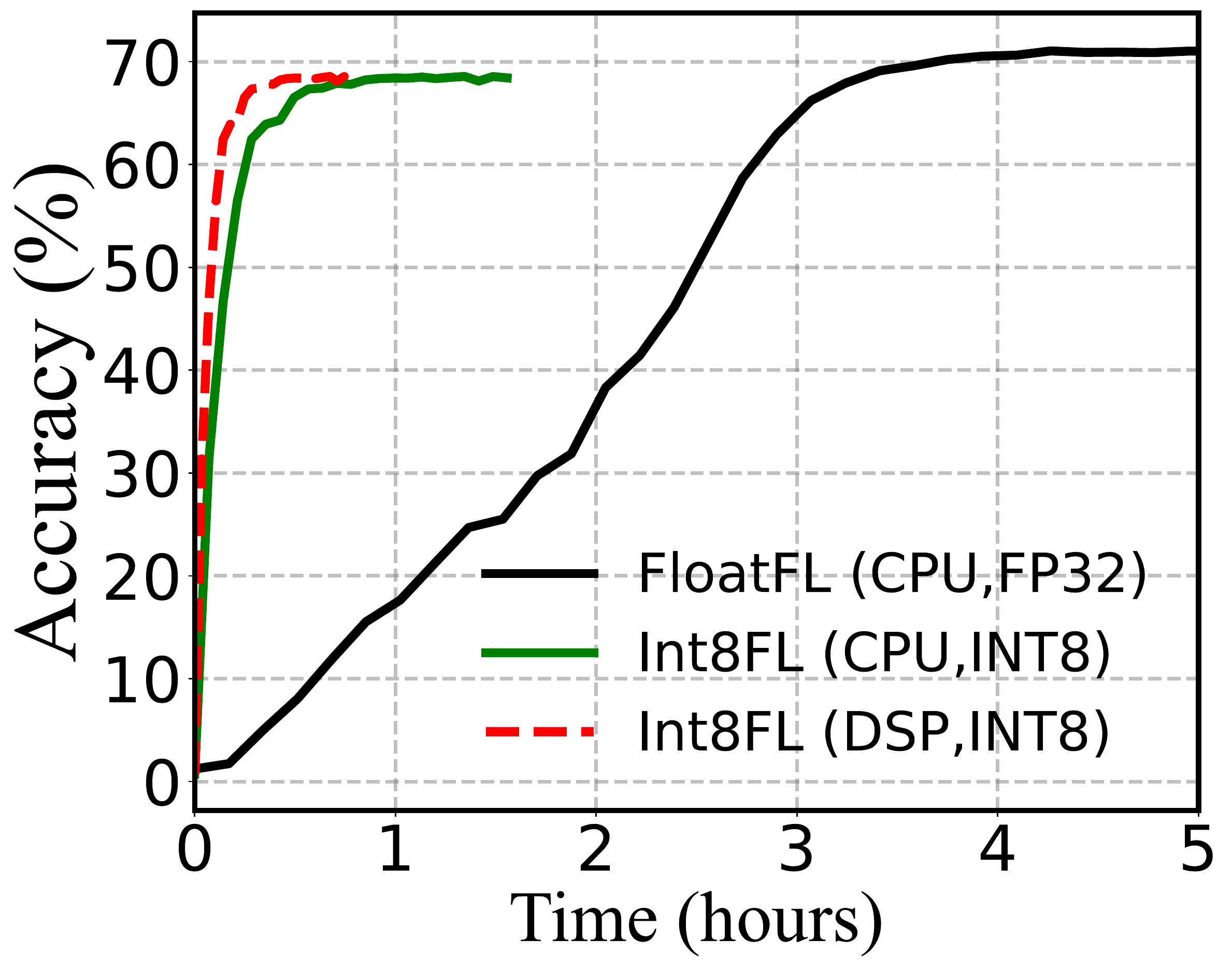}  } 
	\vspace{-0.5cm}
	\caption{The convergence accuracy across clock time under single device and federated scenarios.} 
	\label{fig-End-to-end}
	\vspace{-0.5cm}
\end{figure}

\begin{table}[t]
	\scriptsize
	\setlength{\tabcolsep}{1mm}{
	\begin{tabular}{|c|c|l|r|rrr|}
		\hline
		\multirow{2}{*}{\textbf{Dataset}} & \multirow{2}{*}{\textbf{Model}} & \multicolumn{1}{c|}{\multirow{2}{*}{\textbf{Methods}}} & \multicolumn{1}{l|}{\multirow{2}{*}{\textbf{Acc.}}} & \multicolumn{3}{c|}{\textbf{Training Cost to Convergence}} \\ \cline{5-7} 
		&  & \multicolumn{1}{c|}{} & \multicolumn{1}{l|}{} & \multicolumn{1}{l|}{\textbf{\begin{tabular}[c]{@{}l@{}}Round\\ number\end{tabular}}} & \multicolumn{1}{l|}{\textbf{\begin{tabular}[c]{@{}l@{}}Clock\\ Hours\end{tabular}}} & \multicolumn{1}{l|}{\textbf{\begin{tabular}[c]{@{}l@{}}Energy\\ (WH)\end{tabular}}} \\ \hline\hline
		\multirow{3}{*}{\begin{tabular}[c]{@{}c@{}}Centralized\\ CIFAR-10\end{tabular}} & \multirow{3}{*}{VGG11} & MNN-FP32 & 89.87\% & \multicolumn{1}{r|}{150} & \multicolumn{1}{r|}{29.13} & 187.01 \\ \cline{3-7} 
		&  & MNN-INT8 & 87.17\% & \multicolumn{1}{r|}{150} & \multicolumn{1}{r|}{24.77} & 153.33 \\ \cline{3-7} 
		&  & Ours & 87.17\% & \multicolumn{1}{r|}{150} & \multicolumn{1}{r|}{7.50} & 31.39 \\ \hline\hline
		
		\multicolumn{1}{|c|}{\multirow{3}{*}{\begin{tabular}[c]{@{}c@{}}Centralized\\ CIFAR-10 \end{tabular}}} &  \multicolumn{1}{l|}{\multirow{3}{*}{ResNet18}} & MNN-FP32 & 92.49\% & \multicolumn{1}{r|}{150} & \multicolumn{1}{r|}{223.55} & 1,435.19  \\ \cline{3-7} 
		\multicolumn{1}{|l|}{} & \multicolumn{1}{l|}{} & MNN-INT8 & 90.62\% & \multicolumn{1}{r|}{150} & \multicolumn{1}{r|}{135.71} & 840.04 \\ \cline{3-7} 
		\multicolumn{1}{|l|}{} & \multicolumn{1}{l|}{} & Ours & 90.62\% & \multicolumn{1}{r|}{150} & \multicolumn{1}{r|}{35.68} & 149.32  \\ \hline\hline
		
			\multicolumn{1}{|c|}{\multirow{3}{*}{\begin{tabular}[c]{@{}c@{}}Federated\\ FEMNIST \end{tabular}}} & \multicolumn{1}{l|}{\multirow{3}{*}{LeNet}} & MNN-FP32 & 84.18\% & \multicolumn{1}{r|}{990} & \multicolumn{1}{r|}{0.97} & 0.00057 \\ \cline{3-7} 
	\multicolumn{1}{|l|}{} & \multicolumn{1}{l|}{} & MNN-INT8 & 82.04\% & \multicolumn{1}{r|}{4,960} & \multicolumn{1}{r|}{0.39} & 0.00029 \\ \cline{3-7} 
	\multicolumn{1}{|l|}{} & \multicolumn{1}{l|}{} & Ours & 82.04\% & \multicolumn{1}{r|}{4,960} & \multicolumn{1}{r|}{0.19} & 0.00007 \\ \hline\hline
		
		\multicolumn{1}{|c|}{\multirow{3}{*}{\begin{tabular}[c]{@{}c@{}}Federated\\ CIFAR-100\end{tabular}}} & \multicolumn{1}{l|}{\multirow{3}{*}{VGG16}} & MNN-FP32 & 71.15\% & \multicolumn{1}{r|}{1,960} & \multicolumn{1}{r|}{8.35} & 2.74 \\ \cline{3-7} 
		\multicolumn{1}{|l|}{} & \multicolumn{1}{l|}{} & MNN-INT8 & 68.42\% & \multicolumn{1}{r|}{2,200} & \multicolumn{1}{r|}{1.56} & 1.26 \\ \cline{3-7} 
		\multicolumn{1}{|l|}{} & \multicolumn{1}{l|}{} & Ours & 68.42\% & \multicolumn{1}{r|}{2,200} & \multicolumn{1}{r|}{0.78} & 0.21 \\ \hline
	\end{tabular} }
\caption{A summary of end-to-end training cost till convergence under different training scenarios.}
\label{tab-end-to-end}
\vspace{-0.7cm}
\end{table}

We now demonstrate that \sys is able to significantly accelerate the model convergence while guaranteeing the model accuracy in end-to-end training experiments.
We focus on the time-to-accuracy metric~\cite{wang2020niti,yang2020training,zhong2020towards,zhang2020fixed,zhu2020towards}.

\noindent \textbf{Learning on a single device} is the case when all training data resides in a single device.
We train VGG16 and ResNet18 with training set CIFAR-10 on XiaoMI 11 Pro  and verify the accuracy after each epoch.
We fix the CPU frequency to the max value and 10 minutes sleep after training 10 epochs to avoid shutting down due to overheating.
Note that sleep time does not count in the time-to-accuracy.

As illustrated in Figure~\ref{fig-End-to-end}(a) and (b) and summarized in the first two rows of Table~\ref{tab-end-to-end}, the convergence accuracy of \sys is only 1.9-2.7\% lower than training with FP32.
This accuracy drop is consistent with the numbers reported by the original algorithm paper~\cite{wang2020niti}, and are generally acceptable by the relevant ML community~\cite{wang2020niti,zhou2021octo,wu2016binarized,lin2015neural,rastegari2016xnor, zhou2018adaptive,lin2017towards,courbariaux2015binaryconnect,jacob2018quantization,zhou2016dorefa,banner2018scalable,chen2017fxpnet, wu2018training, zhu2016trained,wang2018training,zhu2020towards,zhang2020fixed,yang2020training,zhong2020towards}.
However, it only takes 5.06-6.27$\times$ less time and 5.96-9.62$\times$ less energy consumption for \sys to the convergence accuracy  (87.17\% and 90.40\%) compared with MNN-FP32.
Compared to MNN-INT8, \sys converges to the same accuracy as they both use NITI training algorithm, but takes 3.55$\times$ less time and 5.46$\times$ less energy consumption on average.

\noindent \textbf{Cross-device federated learning} is another killer use case that allows many devices to collaboratively train a model without giving away their training data.
In our experiments, we use a popular FL simulation platform~\cite{yang2021characterizing} and plug in the tested on-device training performance of \sys and baselines into the platform.
Both FloatFL and Int8FL use  the traditional FL protocol \emph{FedAvg}.
For a fair comparison, we set the number of the local epoch as 1 for all experiments. 
We use FEMNIST~\cite{reddi2020adaptive} and CIFAR-100~\cite{krizhevsky2009learning} as the testing datasets, and follow prior work~\cite{he2020fedml} to partition them into non-IID distribution.

As demonstrated in Figure~\ref{fig-End-to-end}(c) and (d) and the last three rows of Table~\ref{tab-end-to-end}, the accuracy of \sys is 2.14\% and 2.73\% lower than FLoatFL on FEMNIST and CIFAR-100, respectively.
However, it only takes 19.58\% and 9.3\%  of clock time to converge (i.e., 5.26$\times$ and 10.75$\times$ speedup) with \sys, respectively.
Such tremendous improvement over FloatFL protocol comes from both reduced on-device training time and communication time of applying INT8-based training.
As Table~\ref{tab-end-to-end} also shows that \sys reduces 8.14$\times$, and 13.1$\times$ energy consumption of single client, respectively.

\noindent  \textbf{Thermal impacts} 
\begin{figure}[t]
	\centering
	\subfigure[\textbf{XiaoMI 11 Pro}]{ \includegraphics[width=0.22\textwidth]{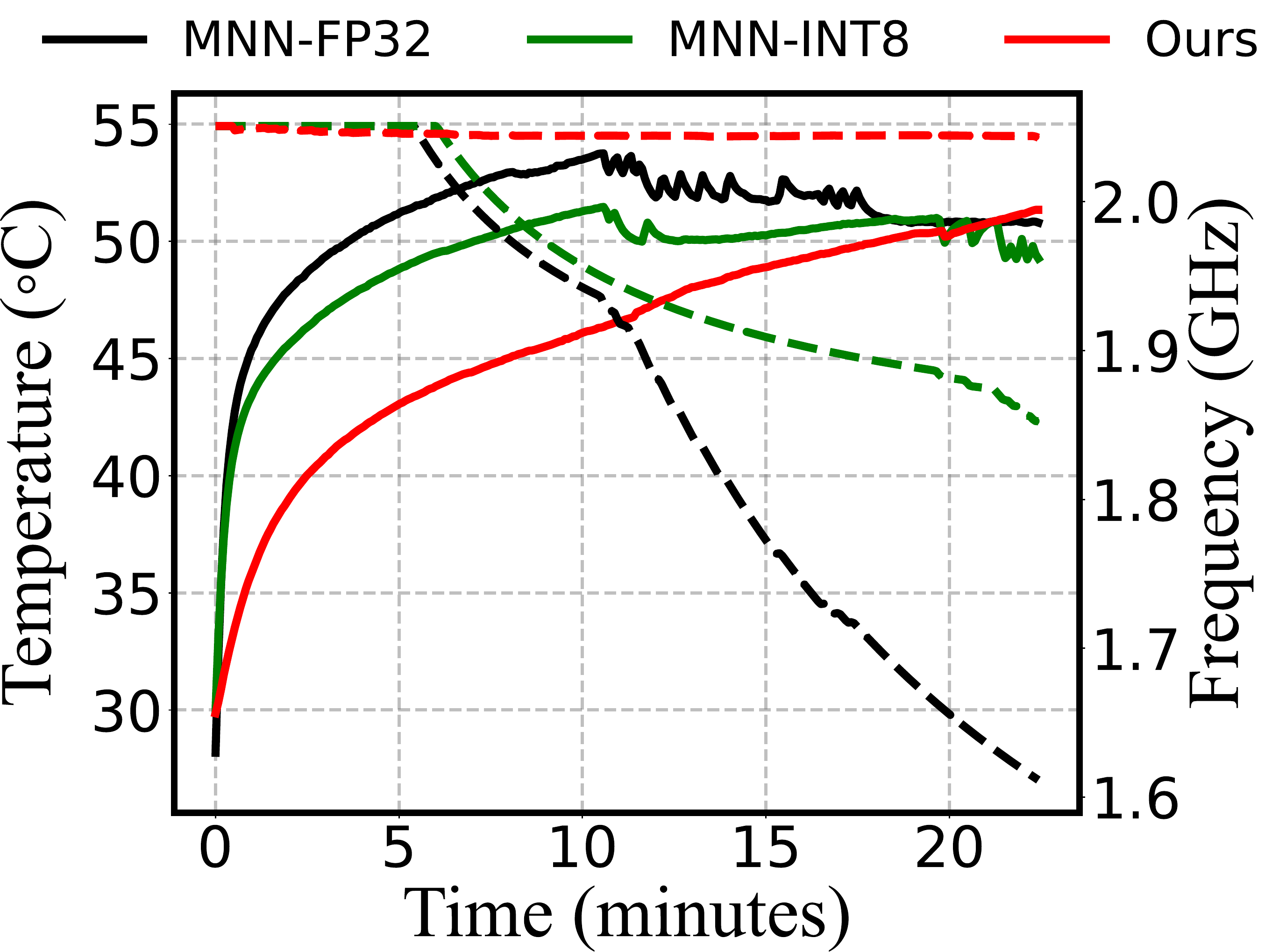} }
	\subfigure[\textbf{XiaoMI 10}]{ \includegraphics[width=0.22\textwidth]{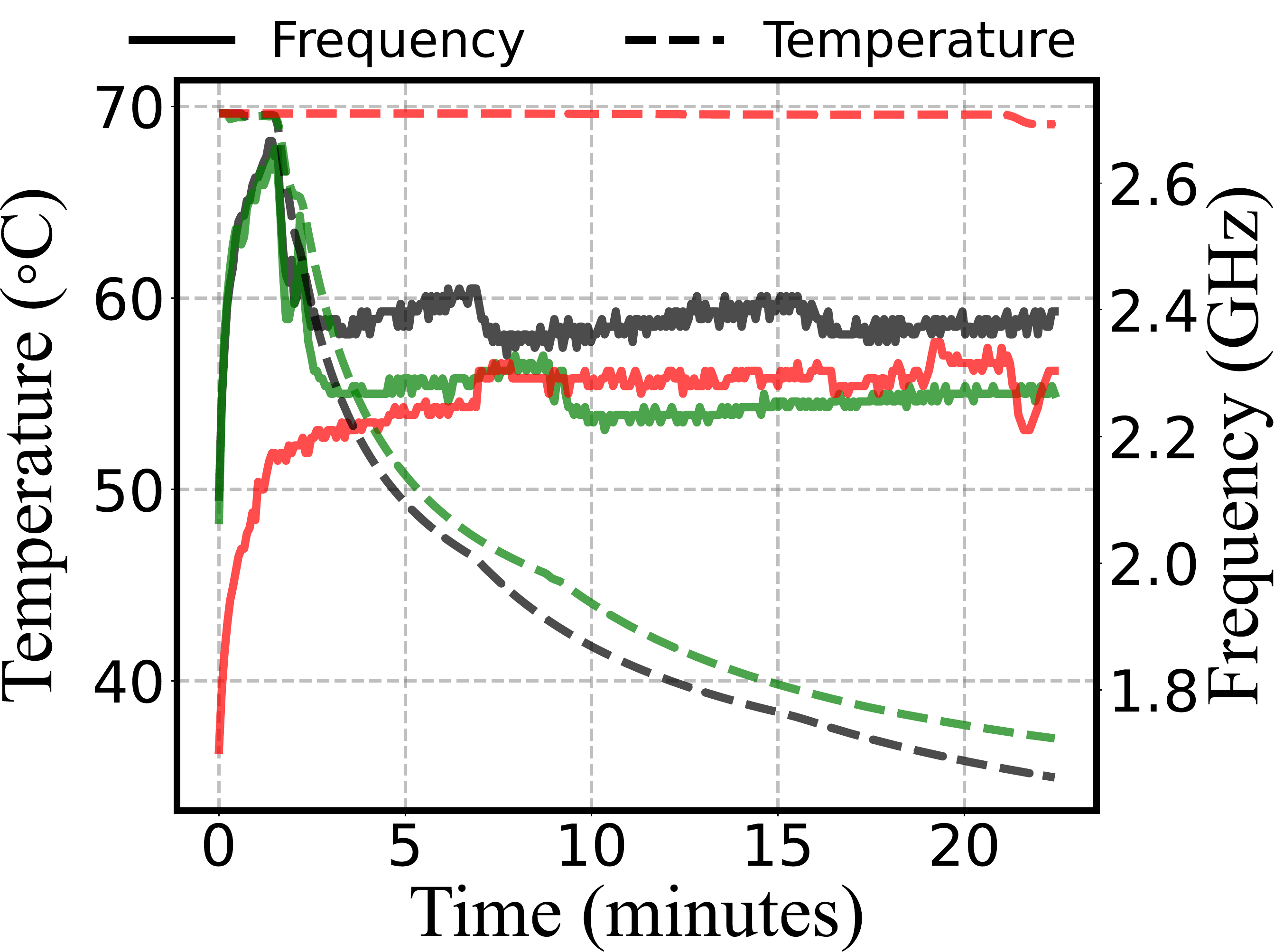} }
	\vspace{-0.5cm}
	\caption{The temperature and CPU frequency dynamics during on-device training on different devices.} 
	\label{fig-thermal}
	\vspace{-0.4cm}
\end{figure}
To reach a usable accuracy, the on-device training often takes a substantial amount of time \cite{bonawitz2019towards}, e.g., minutes for each round of federated learning or even hours for continuous local transfer learning \cite{deeptype}.
Such a long duration of intensive computation may lead to thermal issues and, therefore, the CPU frequency change due to DVFS. 
Thus we investigate the thermal dynamics of on-device training on two devices and illustrate the results in Figure~\ref{fig-thermal}.
On both tested devices, we observe the temperature rising and the CPU underclocking, but the trend for \sys is much milder than the other two baselines.
Using MMN-FP32 and XiaoMI 10 as an example, the device temperature rises sharply from 29°C to 68°C in 2 minutes, and the CPU frequency drops from 2.8GHz to 1.3GHz.
On the other hand, it takes about 18 minutes for \sys to raise the temperature from 29°C to 58°C which leads to almost no CPU underclocking.
This is because DSP frequency is 4$\times$ lower than CPU and is designed for low-power scenarios.



\subsection{Ablation Study}
\begin{figure}[t]
	\centering
	 \includegraphics[width=0.45\textwidth]{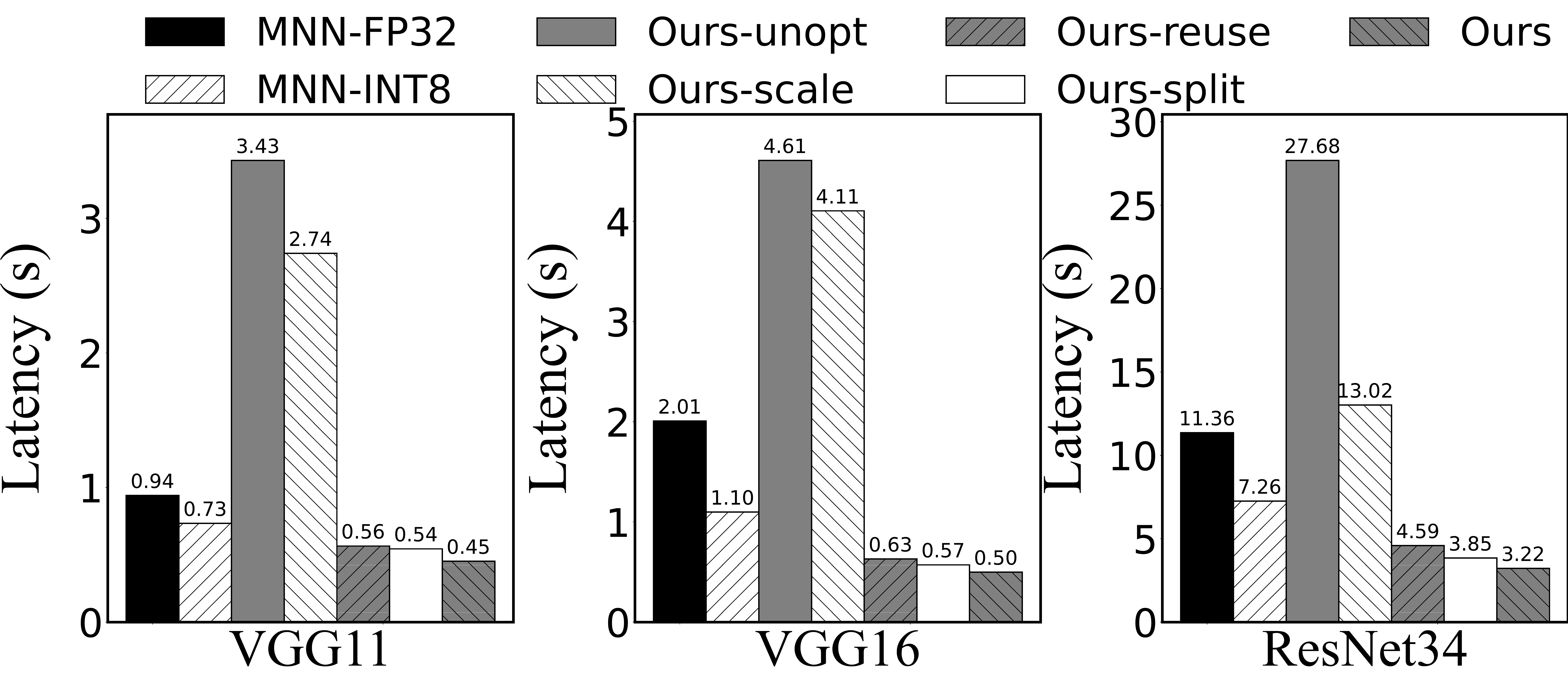} 
	 \vspace{-0.3cm}
	\caption{Ablation study of \sys.} 
	\label{fig-ablation}
	\vspace{-0.5cm}
\end{figure}
We further conduct a breakdown analysis of the benefit brought by \sys's each technique. 
The experiments are performed with VGG11, VGG16 and ResNet34 models on XiaoMI 10.
The results are illustrated in Figure~\ref{fig-ablation}.

We observe that all techniques have non-trivial contribution to the improvement.
For instance, the per-bach training time for ResNet34 model is 27.68s without any optimizations.
When the self-adaptive rescaling is applied, the latency reduces to 13.02s.
Adding the compute subgraph reusing technique further decreases the latency to 4.54s.
The other two techniques, i.e., batch splitting and CPU-DSP parallel execution, also add to 22.9\% and 19.5\% lower latency, respectively.
Since the 4 key techniques optimize the training cost from different aspects, they can well orchestrate to provide accumulative optimization.
Besides, the profits of batch splitting for VGG11 is rather small compared with other two models.
That is because the workload of VGG11 model is smaller so we do not observe high cache pressure that motivates the batch splitting technique.

\subsection{Fitting to Various Training Algorithms}
\begin{figure}[t]
	\centering
	\subfigure[\textbf{Latency}]{
	\includegraphics[width=0.45\textwidth]{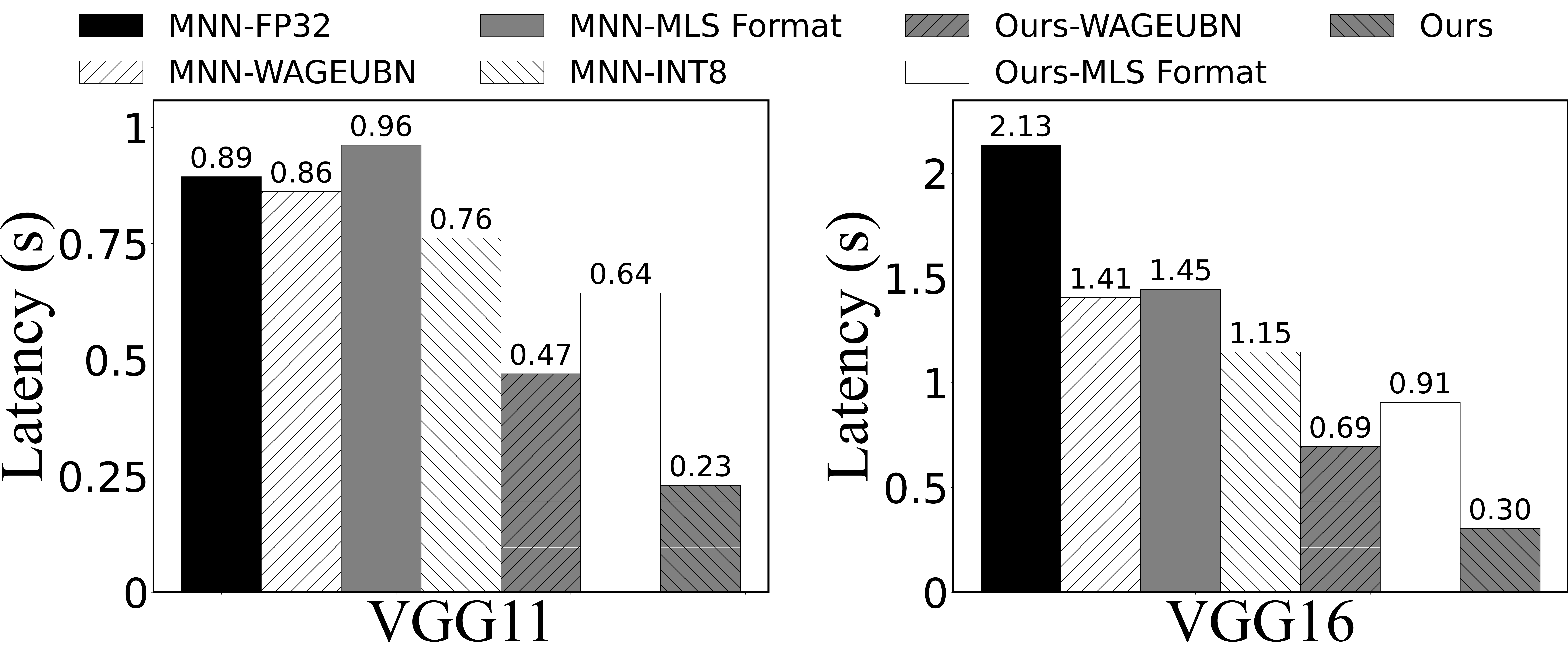} }
	\subfigure[\textbf{Energy consumption}]{
	\includegraphics[width=0.45\textwidth]{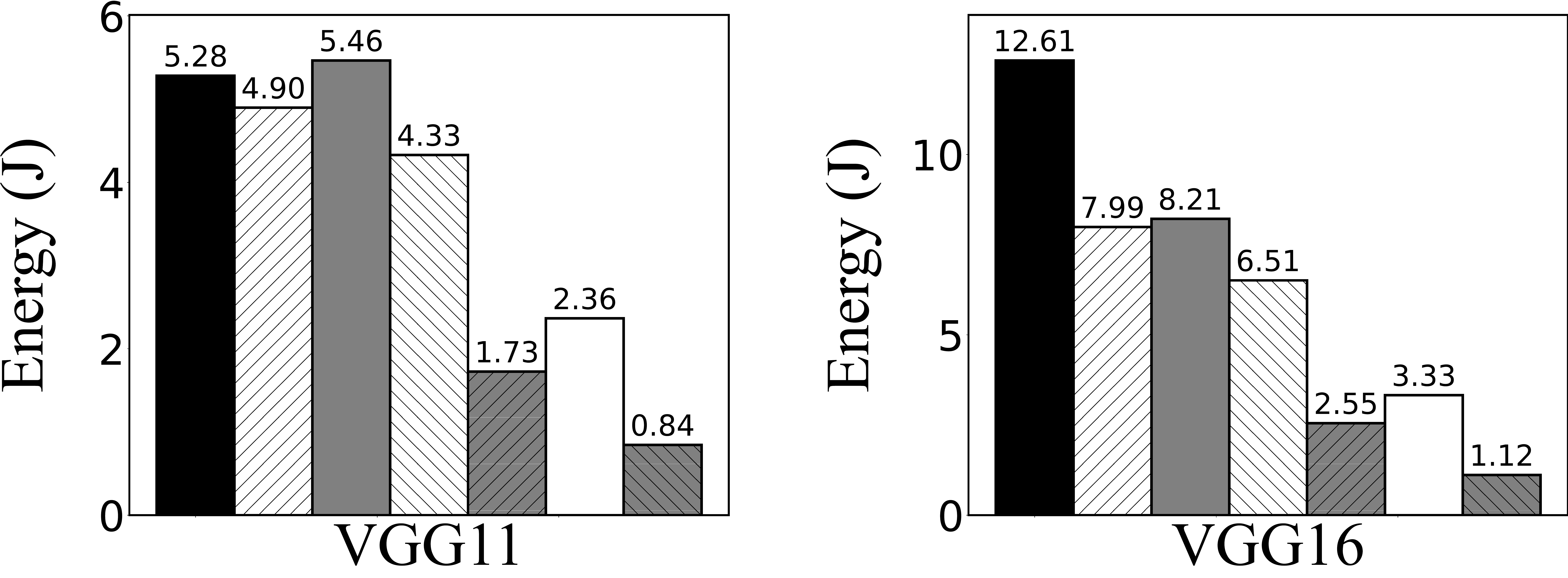} }
\vspace{-0.5cm}
	\caption{\sys's performane using different mixed-precision training algorithm. 
		"MNN-*" are baselines runing on CPU; while "Ours-*" are runing on DSP using \sys.} 
	\label{fig-low-precision}
	\vspace{-0.5cm}
\end{figure}

Recall that \sys is an underlying independent framework that supports different kinds of mixed-precision training algorithms.
Therefore we also test \sys with different training algorithms: NITI~\cite{wang2020niti} (default), MLS Format \cite{zhong2020towards}, and WAGEUBN \cite{yang2020training}.
The experiments are performed with VGG11 and VGG16 on XiaoMI 11 Pro.

As shown in Figure~\ref{fig-low-precision}, the training speed of three mixed-precision algorithms are all faster than that of MNN-FP32 (3.08$\times$, 2.34$\times$, and 7.1$\times$ speedup).
The energy consumption improvement of \sys is more significant, i.e., 3.64$\times$  and 6.85$\times$ average reduction for VGG11 and VGG16, respectively.
That is because \sys's key techniques are aimed at solving the generic challenges to support different kinds of mixed-precision training algorithms. 
When comparing the same mixed-precision training algorithms running on CPU and DSP, \sys is still 2.04$\times$, 1.59$\times$, and 3.83$\times$ faster, respectively.
Such benefit comes from \sys's effective DSP offloading.
Besides, among different training algorithms, NITI has the lowest training latency because it maximizes the usage of INT8 in its data flow. 

	\section{Related Work}\label{sec:related}

\noindent \textbf{Mixed-precision DNN training}
Recently, many mixed-precision DNN training algorithms have been proposed to reduce the training cost \cite{wang2020niti,zhou2021octo,wu2016binarized,lin2015neural,rastegari2016xnor, zhou2018adaptive,lin2017towards,courbariaux2015binaryconnect,jacob2018quantization,zhou2016dorefa,banner2018scalable,chen2017fxpnet, wu2018training, zhu2016trained}. 
The key idea is to replace the default numerical format of FP32 with lower precision for activations and/or weights, e.g., FP16, INT8, or even BOOL type.
Octo \cite{zhou2021octo} further proposes an INT8-based training algorithm and also builts a system for edge GPUs.
Instead of contributing new mixed-precision training algorithms, \sys is designed as an generic, underlying system to efficiently support those algorithms.

\noindent \textbf{On-chip DNN offloading}
has been studied to enable faster DNN inference on heterogeneous mobile processors like GPU and DSP.
Most of them focus on how to partition and schedule the workloads on different processors, such as  intra-layer \cite{kim2019mulayer,zeng2021energy}, inter-layer \cite{ha2021accelerating}, block-layer level \cite{lane2016deepx,han2019mosaic}, and model level \cite{zhang2020mobipose,lee2019mobisr,georgiev2014dsp}.
However, they only focus on DNN inference.
Besides ML tasks, GPU/DSP are also leveraged for signal, sound and image processing \cite{lane2015deepear,georgiev2017accelerating,rana2016opportunistic,liu2014imashup}.
\sys is motivated by those efforts, and is the first framework to support the offloading of training workloads to mobile DSP.

\noindent \textbf{DNN inference optimizations}
Besides on-chip DNN offloading, there have many other research efforts in improving the DNN inference performance on devices.
For instance, some apply structured pruning techniques to trade off latency and model accuracy \cite{han2021legodnn,fang2018nestdnn,han2016mcdnn}.
Some improve the low-level kernel implementation through a compiler, core scheduling, etc~\cite{wang2021asymo,chen2018tvm,liu2019optimizing}.
Some of them optimize the DNN inference in specific scenarios \cite{huynh2017deepmon,xu2018deepcache,yeo2020nemo,zhang2020mobipose}.
\sys is inspired by those work, yet focuses on training instead of inference.
As previously discussed, DNN training faces many unique challenges as compared to inference, so \sys contributes novel techniques in addressing those challenges.

\noindent \textbf{Federated learning}
is an emerging machine learning paradigm~\cite{konevcny2016federatedlearning,mcmahan2017communication,niu2020billion,mo2021ppfl} that
is built atop on-device training and requires many clients to collaboratively train a DNN model.
The communication bottleneck seriously affects the system efficiency and model accuracy \cite{kairouz2021advances}.
Therefore, prior work mostly focus on model compression technique \cite{jhunjhunwala2021adaptive,reisizadeh2020fedpaq,shlezinger2020federated,tang2018communication,li2021hermes} to reduce communication traffic.
Others 
As an underlying framework, \sys is orthogonal and compatible with those algorithm-level optimizations.
\section{Conclusions}\label{sec:conclusions}
In this paper, we have proposed  \sys, the first system that enables highly resource-efficient on-device training by orchestrating the mixed-precision training with on-chip DSP offloading.
\sys incorporated novel techniques such as self-adaptive rescaling and CPU-DSP co-scheduling to fully unleash the power of DSP.
We conducted extensive experiments to evaluate \sys.

	\bibliographystyle{plain}
	\bibliography{ref-mwx}

\end{document}